\documentclass[preprint,12pt]{elsarticle}
\usepackage{natbib}
\setcitestyle{authoryear,round}
\usepackage{graphicx}
\usepackage{soul}
\soulregister\cite7

\usepackage{amssymb}

\usepackage{graphicx}
\usepackage{subfigure}

\usepackage{url}

\journal{Planetary and Space Science}
\begin{document}

\begin{frontmatter}



\title{Photometric study for near-Earth asteroid (155140)~2005~UD}


\author[author1,author2,author3]{Huang, J.-N.}
\author[author4,author5]{Muinonen, K.}
\author[author6]{Chen, T.}
\author[author1,author2,author3]{Wang, X.-B.\corref{cor1}}
\cortext[cor1]{Corresponding author: Xiaobin Wang\\wangxb@ynao.ac.cn, huangjianing@ynao.ac.cn}

\address[author1]{Yunnan Observatories, Chinese Academy of Sciences, Kunming, 650216, China}

\address[author2]{School of Astronomy and Space Sciences, University of Chinese Academy of Sciences, Beijing, 100049, China}

\address[author3]{Key Laboratory for Structure and Evolution of Celestial Objects, CAS, Kunming, 650216, China}

\address[author4]{Department of Physics, University of Helsinki, Gustaf H\"allstr\"omin katu 2, P.O. Box 64, FI-00014 U. Helsinki, Finland}

\address[author5]{Finnish Geospatial Research Institute, Geodeetinrinne 2, FI-02430 Masala, Finland}

\address[author6]{Corona borealis Observatiories, Ali, China }

\begin{abstract}
  The Apollo-type near-Earth asteroid (155140)~2005~UD is thought to
  be a member of the Phaethon-Geminid meteor stream Complex (PGC). Its
  basic physical parameters are important for unveiling its origin and
  its relationship to the other PGC members as well as to the Geminid
  stream. Adopting the Lommel-Seeliger ellipsoid method and
  $H,G_1,G_2$ phase function, we carry out spin, shape, and phase
  curve inversion using the photometric data of 2005~UD. The data
  consists of 11 new lightcurves, 3 lightcurves downloaded from the
  Minor Planet Center, and 166 sparse data points downloaded from the
  Zwicky Transient Facility database.  As a result, we derive the pole
  solution of ($285^\circ.8^{+1.1}_{-5.3}$,
  $ -25^\circ.8^{+5.3}_{-12.5}$) in the ecliptic frame of J2000.0 with
  the rotational period of $5.2340$ h. The corresponding triaxial
  shape (semiaxes $a>b>c$) is estimated as $b/a= 0.76^{+0.01}_{-0.01}$
  and $c/a=0.40^{+0.03}_{-0.01}$. Using the calibrated photometric data
  of 2005~UD, the $H,G_1,G_2$ parameters are estimated as
  $17.19^{+0.10}_{-0.09}$ mag, $0.573^{+0.088}_{-0.069}$, and
  $0.004^{+0.020}_{-0.021}$, respectively. Correspondingly, the phase
  integral $q$, photometric phase coefficient $k$, and the enhancement
  factor $\zeta$ are 0.2447, -1.9011, and 0.7344. From the values of
  $G_1$ and $G_2$, 2005~UD is likely to be a C-type asteroid. We
  estimate the equivalent diameter of 2005~UD from the new $H$-value:
  it is 1.30~km using the new geometric albedo of 0.14.
\end{abstract}




\end{frontmatter}


\section{Introduction}

Near-Earth asteroids (NEAs) derive from the main belt of asteroids and
are composed of planetesimal material remaining from the early stages
of the Solar System. They contain important information about the
Solar System's formation and evolution. Thus, the physical study for
NEAs can allow us to know more about the history of the Solar System,
including that of the asteroids. Asteroid 2005~UD has been discovered
by the Catalina Sky Survey in 22 October 2005, and, according to its
orbit, it has been classified as an Apollo-type NEA. Together with
(3200)~Phaethon and 1996~YC, 2005~UD belongs to the Phaethon-Geminid
stream Complex, briefly PGC.\cite{2006A&A...450L..25O} investigated
the orbital evolution of 2005~UD and Phaethon, and suggested the
object may be a split nucleus of Phaethon. As a member of PGC, the
physical properties of 2005~UD should provide information about its
origin and relationship to the other PGC members and to the Geminid
meteors.

Several groups have carried out physical studies of
2005~UD. \cite{2006A&A...450L..25O} inferred that 2005~UD is a km-size
object. {\cite{2006AJ....132.1624J}'s data of 2005~UD showed a
periodic brightness variation of 5.2492~h with an amplitude of
0.4~mag. They also estimated the diameter of 2005~UD to be
$1.3\pm0.1$~km, assuming a geometric albedo of $0.11$ (the geometric
albedo of Phaethon). \cite{2007A&A...466.1153K} determined the
rotation period for 2005~UD, and found variation of color indices with
rotational phase. They derived a rotation period of 5.23~h with a
lightcurve amplitude of $0.44\pm0.02$~mag. Their multi-band
photometric observations suggested that 2005~UD is of F or
B type. Later, \cite{2008AJ....136..881K} suggested that 2005~UD is a
C-type asteroid, based on the color index. \cite{2008AJ....136..881K}
determined the absolute magnitude of 2005~UD to be $17.23\pm0.03$~mag,
and derived a diameter of $1.2\pm0.1$~km, assuming an albedo of
$0.11\pm0.02$. \cite{2013AJ....145..133J} re-determined the surface
color indices of 2005~UD and estimated its absolute magnitude to be
17.08~mag, assuming the $H,G$ phase function slope of $G=0.15$. Based
on the WISE observations, \cite{2019AJ....158...97M} determined the
geometric albedo of 2005~UD to be $0.14\pm0.09$, and derived a diameter
of $1.2\pm0.4$~km. Recently, \cite{2019EPSC...13.1989K} and
\cite{2019MPBu...46..144W} reported a similar rotation period for
2005~UD. \cite{2019EPSC...13.1989K} determined the linear phase-angle
coefficient of 0.043~mag~deg$^{-1}$.} As for the spin and shape
parameters, there is no further information preceding our study.

For understanding the physical properties of 2005~UD, we have carried
out 11 nights of photometric observations in 2018. In addition, we
download the photometric data of 2005~UD from the Zwicky Transient
Facility (ZTF) and Minor Planet Center (MPC) databases. ZTF is a
robotic time-domain astronomical sky survey. The small Solar System
bodies are important targets of the survey.In total, 166 data points
of 2005~UD are downloaded from the ZTF database.

In the article, we present our new photometric observations of 2005~UD
and our results of the photometric analyses of 2005~UD with the
inverse methods based on Lommel-Seeliger ellipsoids
(\cite{2015P&SS..118..227M}) and $H,G_1,G_2$ phase curves
(\cite{2010Icar..209..542M}). Therefore, Sect. 2 shows the observations
and data reductions for the object, Sect. 3 introduces the methods of
lightcurve inversion and phase curve inversion, together with the
results and discussion. In the last section, a summary is presented.

\section{Observations and data reductions}

We obtained 11 nights of photometric observations of 2005~UD in
October 2018 using the 30-cm telescope and a 3326$\times$2504 CCD at
the Corona borealis observatory (code N55). The field of view (FOV) of
the CCD is $28'.4 \times 21'.4$. The data were gathered through C
(Clear), R, or V filters depending on the weather conditions and the
signal-to-noise ratio. During the observations, the weather conditions
were acceptable with good seeing, and the status of the instrument was
fine. Additional observational information is shown in Table~1. During
the photometric observations, sidereal tracking and short exposure
times were utilized. During the observations, the sky-plane motion of
2005~UD was large at about $8''.7\sim3''.5/ min$. Consequently, the
telescope pointing was shifted once or twice in some nights to keep
the object in the FOV.

The photometric images were reduced according to the standard
procedures with the IRAF software. The effects of bias, flat field,
and dark current were corrected first. The cosmic rays in the images
were removed properly. The magnitudes of the celestial objects in the
scientific frames were measured using the Apphot task of IRAF with an
optimum aperture. We tried 3-5 apertures ranging 2.0-2.3 times the
full width at half maximum (FWHM) to find the optimum aperture giving
the minimum dispersion of the lightcurve points.

Some systematic errors in the photometric data, related to atmospheric
extinction and temporal or positional changes of stars in the CCD,
were simulated with the aid of reference stars in the images using the
coarse de-correlation method
(\cite{2006MNRAS.373..799C},\cite{2013RAA....13..593W}) and the SYSREM
method (\cite{2005MNRAS.356.1466T}). The former performs a coarse
initial de-correlation by referencing each star's magnitude to its own
mean, finding small night-to-night and frame-to-frame differences in
the zero point. After the coarse de-correlation, the reduced
magnitudes of celestial objects can be derived (see Eq. 1d below) by
removing the biases due to the objects' mean magnitude in each night
and zero-point in each frame. The latter method simulates the
low-level systematic errors in the reduced magnitudes by those chosen
reference stars. Then those simulated low-level errors, or, say,
patterns in the reduced magnitudes are removed from the reduced
magnitudes of the 2005~UD. The ratio of signal to noise of the
lightcurves (reduced magnitudes in one night) is therefore
enhanced. The time stamp of each observation of the asteroid is
corrected for the light travel time. The distance effects on the
asteroid's magnitude are also corrected by the formula
$-5\log(r\Delta)$.In total, 2206 data points were obtained in 11
nights. For the aim of shape inversion, the relative intensities of
2005~UD are used which are derived by normalizing the mean intensity
of each lightcurve to unity.

\begin{table*}[h]
  \caption{Information on the phototometric observations of near-Earth asteroid 2005~UD.}\label{T:global}
	\centering
	\scalebox{0.7}{
	\begin{tabular}{cccccccccccc}
		\hline
		Date			        	& \textbf{r}  	& \textbf{$\Delta$} 	& \textbf{$\alpha$} 	& \textbf{Mag-V}   & \textbf{Filter}	& \textbf{N} &  \textbf{Data source}&\\
		(UT	)					& \textbf{(au)}    & \textbf{(au)}            & \textbf{(deg)} 	& \textbf{}          & \textbf{}   &\textbf{} & \textbf{} &\\
		\hline
		2018/10/05	 & $1.219$	& $0.245$ 	& $21.301$	& $15.9$ 	 & $C$  &$135$ &$Our data$  \\
		2018/10/06	 & $1.232$	& $0.252$ 	& $17.588$	& $15.9$ 	 & $C$  &$229$	 &$Our data$ \\
		2018/10/07	 & $1.246$  & $0.261$ 	& $14.198$	& $15.9$ 	 & $C$ 	&$263$   &$Our data$ \\
		2018/10/10	 & $1.286$  & $0291$ 	& $5.671$ 	& $15.8$     & $R$  &$36$	 &$Our data$ \\
		2018/10/11	 & $1.299$	& $0.302$ 	& $3.152$ 	& $15.8$     & $V,R$  &$227$  &$Our data$ \\
          2018/10/12	 & $1.312$	& $0.315$ 	& $1.790$ 	& $15.7$ 	 & $C$   &$52$  &$R,Stephens$ \\
		2018/10/12	 & $1.312$	& $0.315$ 	& $0.964$ 	& $15.7$ 	 & $V$   &$204$  &$Our data$  \\
		2018/10/13	 & $1.325$	& $0.327$ 	& $1.726$	& $15.7$ 	 & $V$ 	&$290$  &$Our data$ \\
		2018/10/14	 & $1.338$	& $0.341$  	& $3.685$	& $15.8$     & $V$ &$292$	 &$Our data$ \\
           2018/10/15	 & $1.350$	& $0.354$ 	& $4.642$	& $16.0$ 	 & $C$ 	&$132$   &$R,Stephens$\\
		2018/10/15	 & $1.350$	& $0.354$ 	& $5.414$	& $16.1$ 	 & $R$ 	&$141$ &$Our data$ \\
           2018/10/16	 & $1.363$	& $0.369$ 	& $6.455$	& $16.3$ 	 & $C$ &$116$	 &$R,Stephens$ \\
           2018/10/16	 & $1.363$	& $0.369$ 	& $7.130$	& $16.5$ 	 & $R$ &$185$	 &$Our data$ \\
		2018/10/17	 & $1.375$	& $0.383$ 	& $8.827$	& $16.6$ 	 & $R$ &$208$  &$Our data$\\
          2017/10/27-2019/7/20& $$ &$$ &$$ &$$ &$$ &$166$&$ZTF^{Facility}$\\ 
		\hline
	\end{tabular}}

      \footnotesize{Note that $r$ and $\Delta$ are heliocentric and topocentric distances of the asteroid, $\alpha$ is the solar phase angle, 
        Mag-V is the mean of the observed V-band magnitudes in a night, N is the number of data points, and $ZTF^{Facility}$ refers to 
        IRSA, Spitzer, WISE, Herschel, Planck, SOFIA, IRTF, IRAS, and MSX.}
\end{table*}

\begin{figure}[htbp]
\centering
\subfigure{
\begin{minipage}[t]{0.35\linewidth}
\centering
\includegraphics[width=5cm]{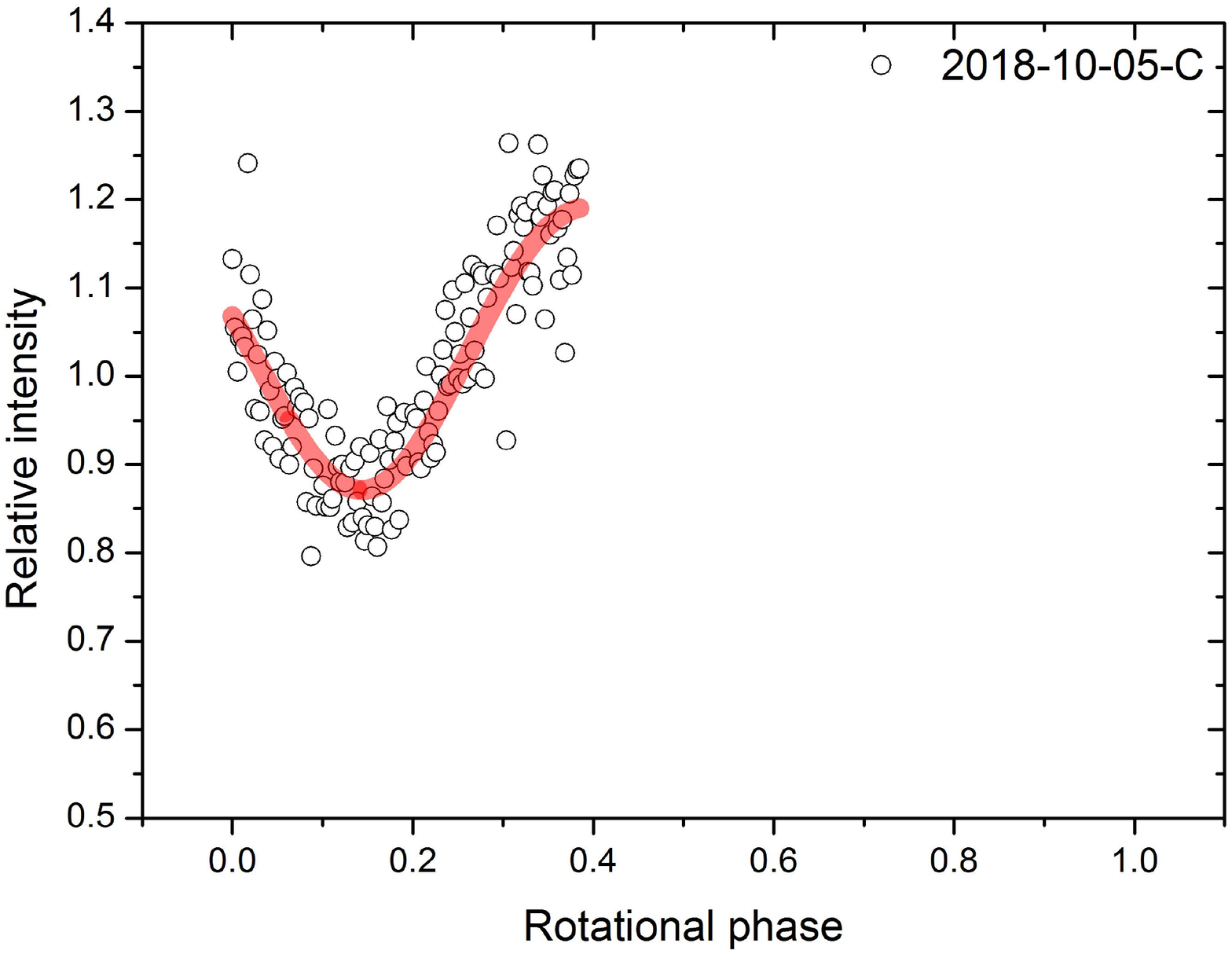}
\end{minipage}%
}%
\subfigure{
\begin{minipage}[t]{0.35\linewidth}
\centering
\includegraphics[width=5cm]{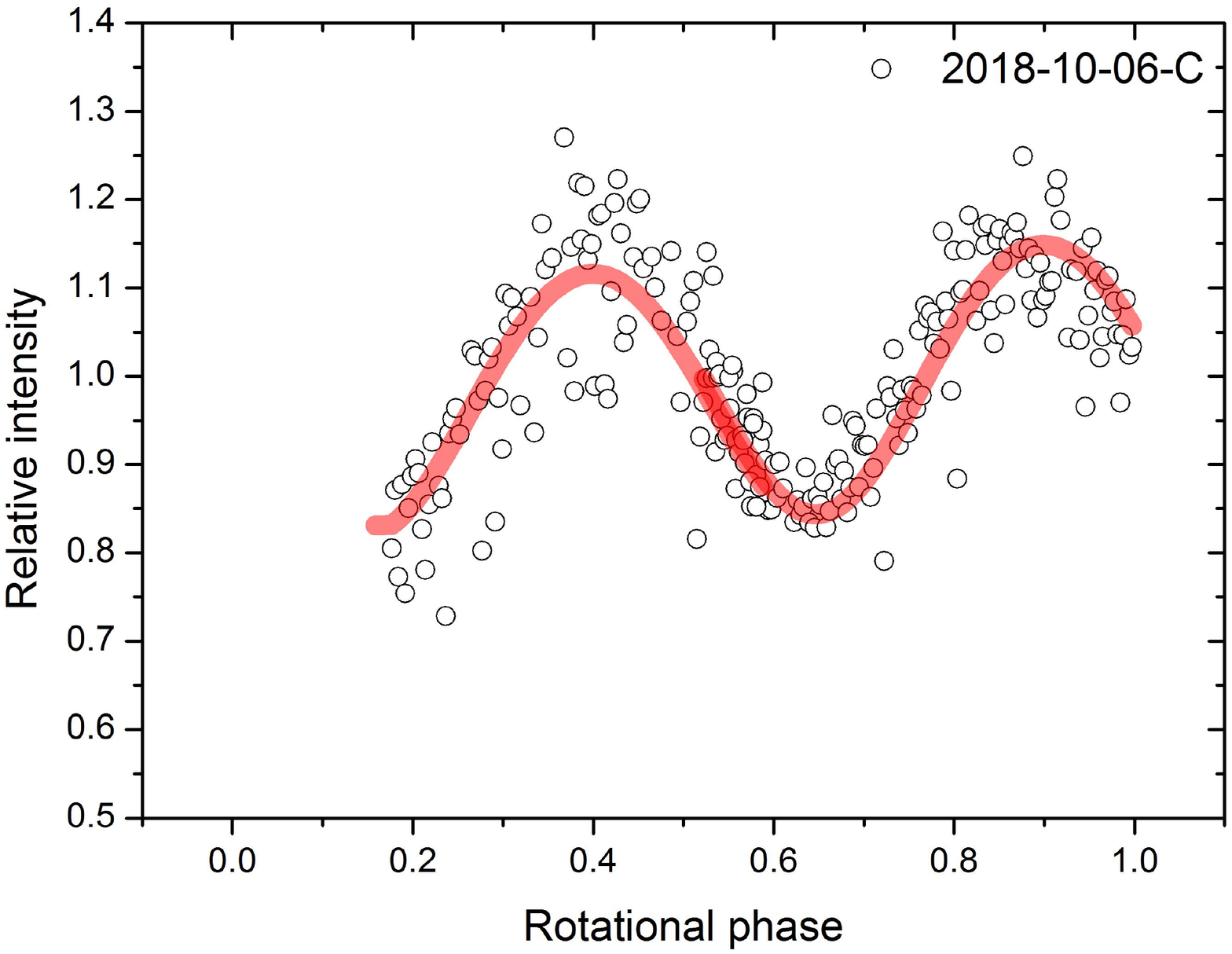}
\end{minipage}%
}%

\subfigure{
\begin{minipage}[t]{0.35\linewidth}
\centering
\includegraphics[width=5cm]{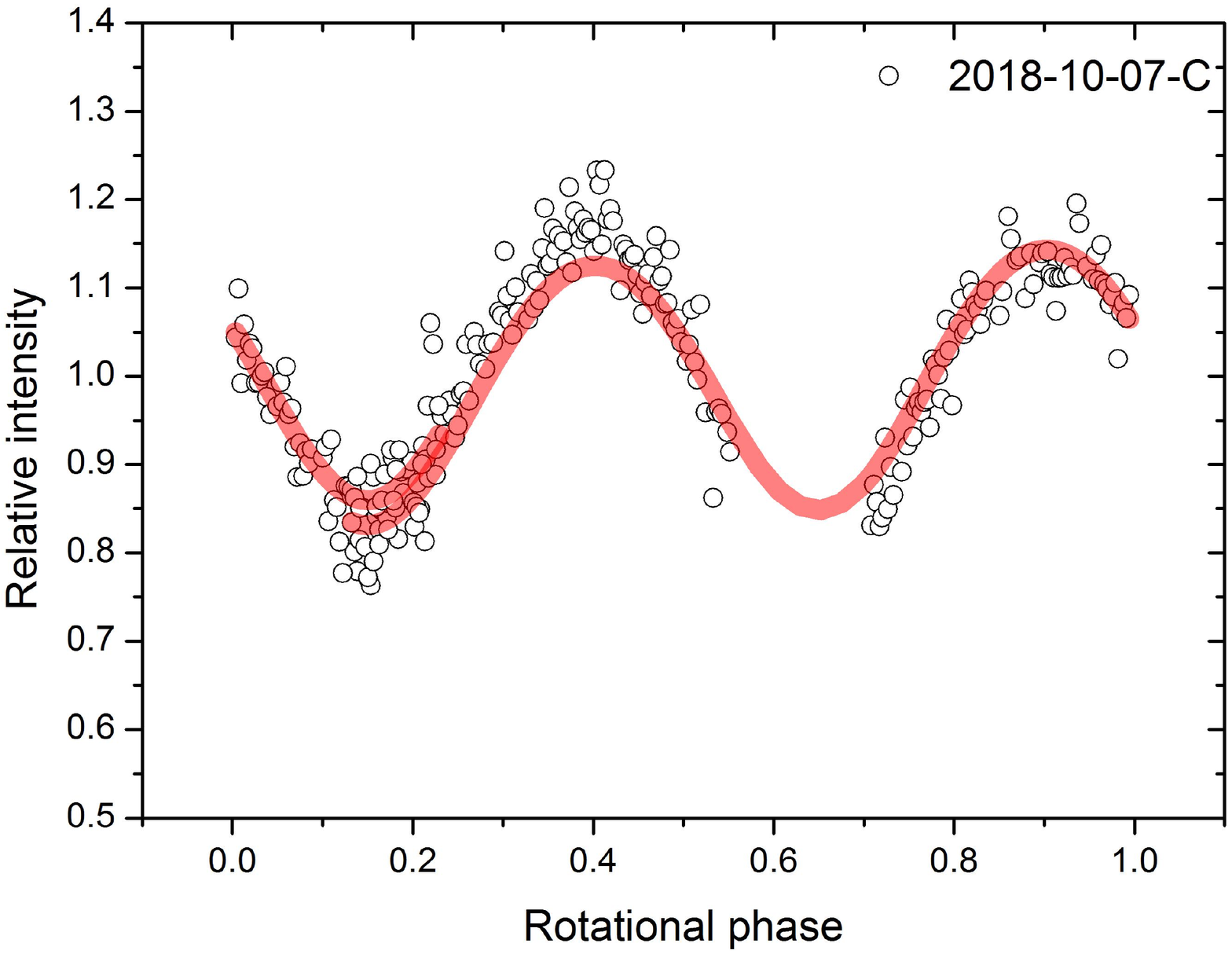}
\end{minipage}%
}%
\subfigure{
\begin{minipage}[t]{0.35\linewidth}
\centering
\includegraphics[width=5cm]{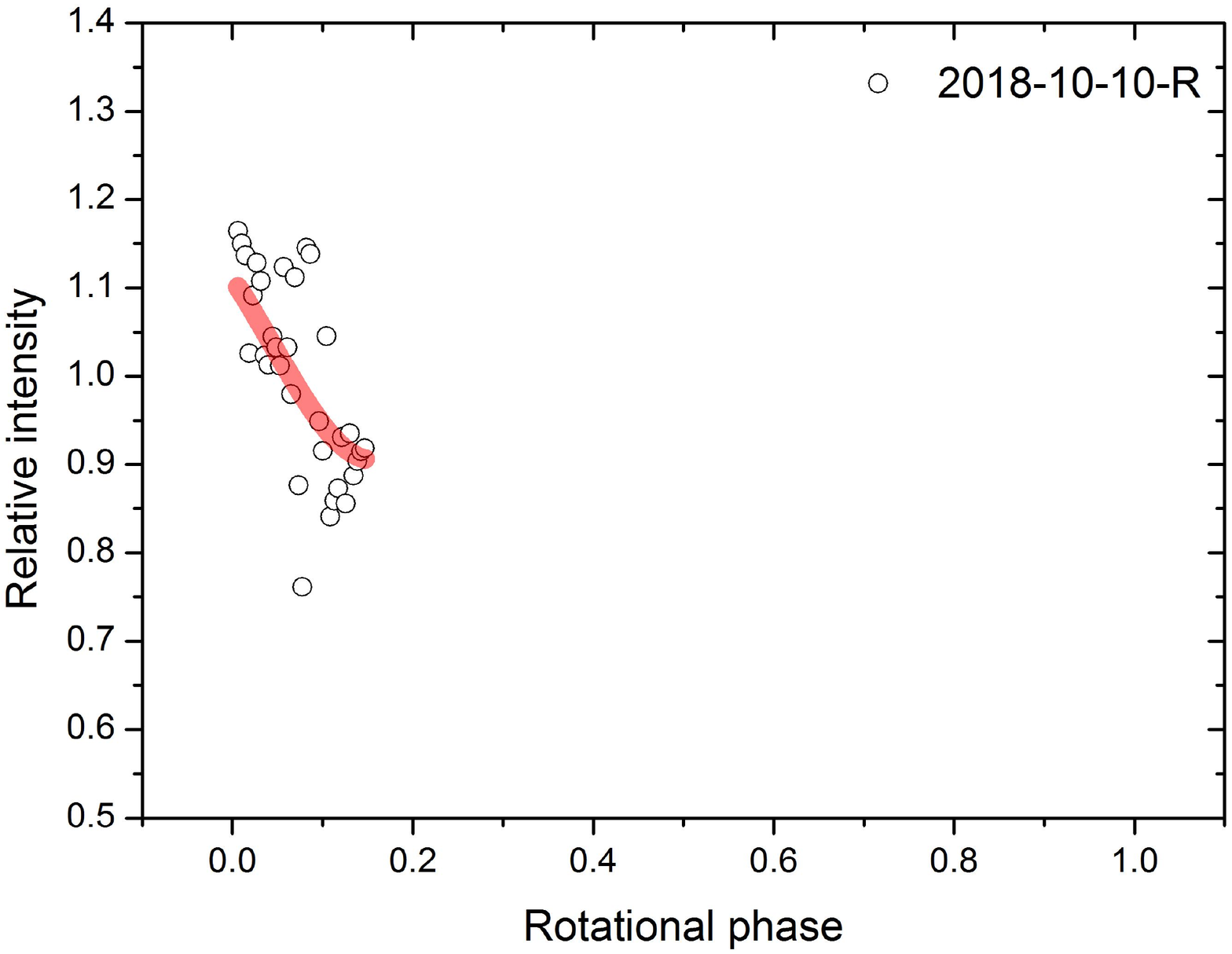}
\end{minipage}%
}%

\subfigure{
\begin{minipage}[t]{0.35\linewidth}
\centering
\includegraphics[width=5cm]{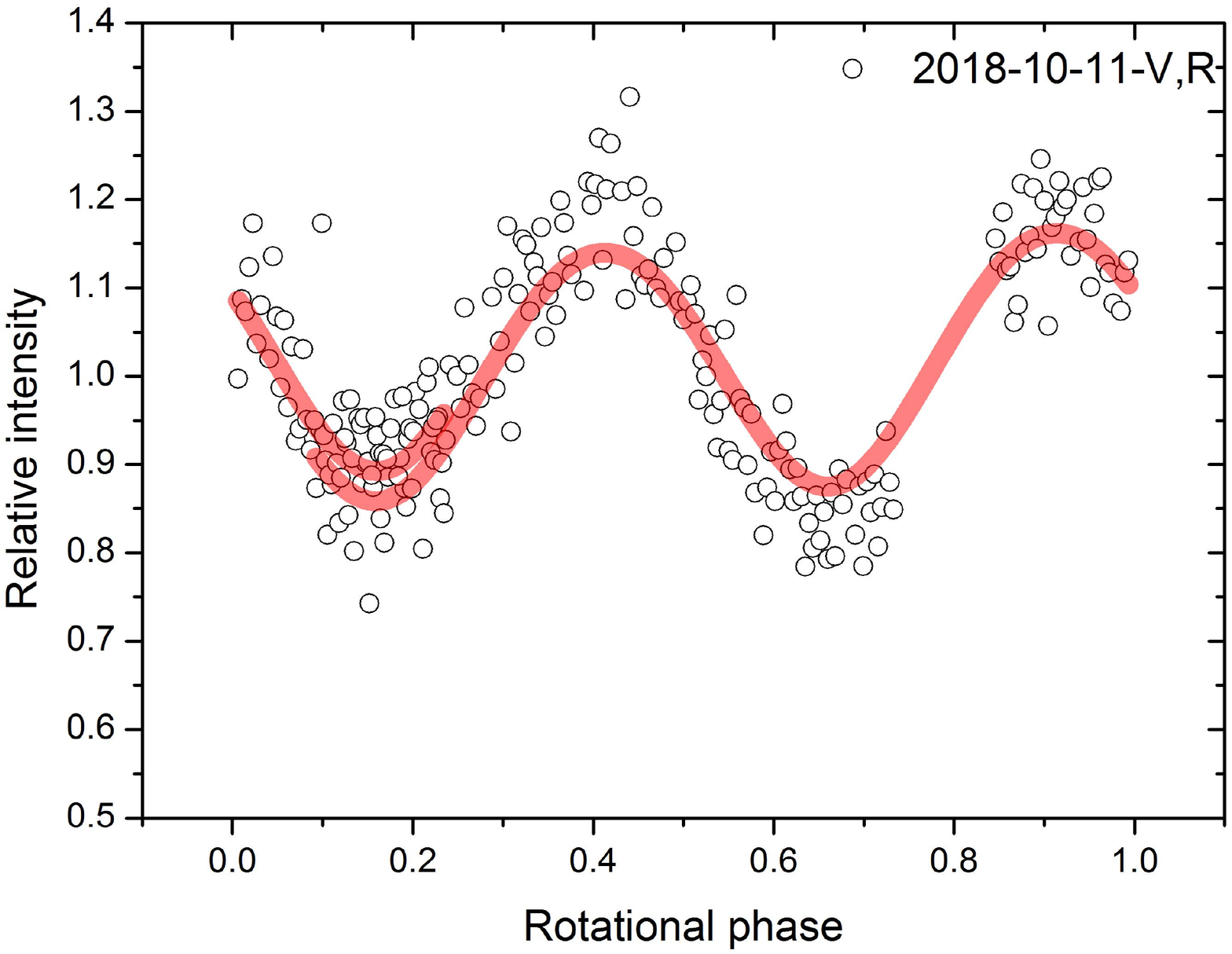}
\end{minipage}%
}%
\subfigure{
\begin{minipage}[t]{0.35\linewidth}
\centering
\includegraphics[width=5cm]{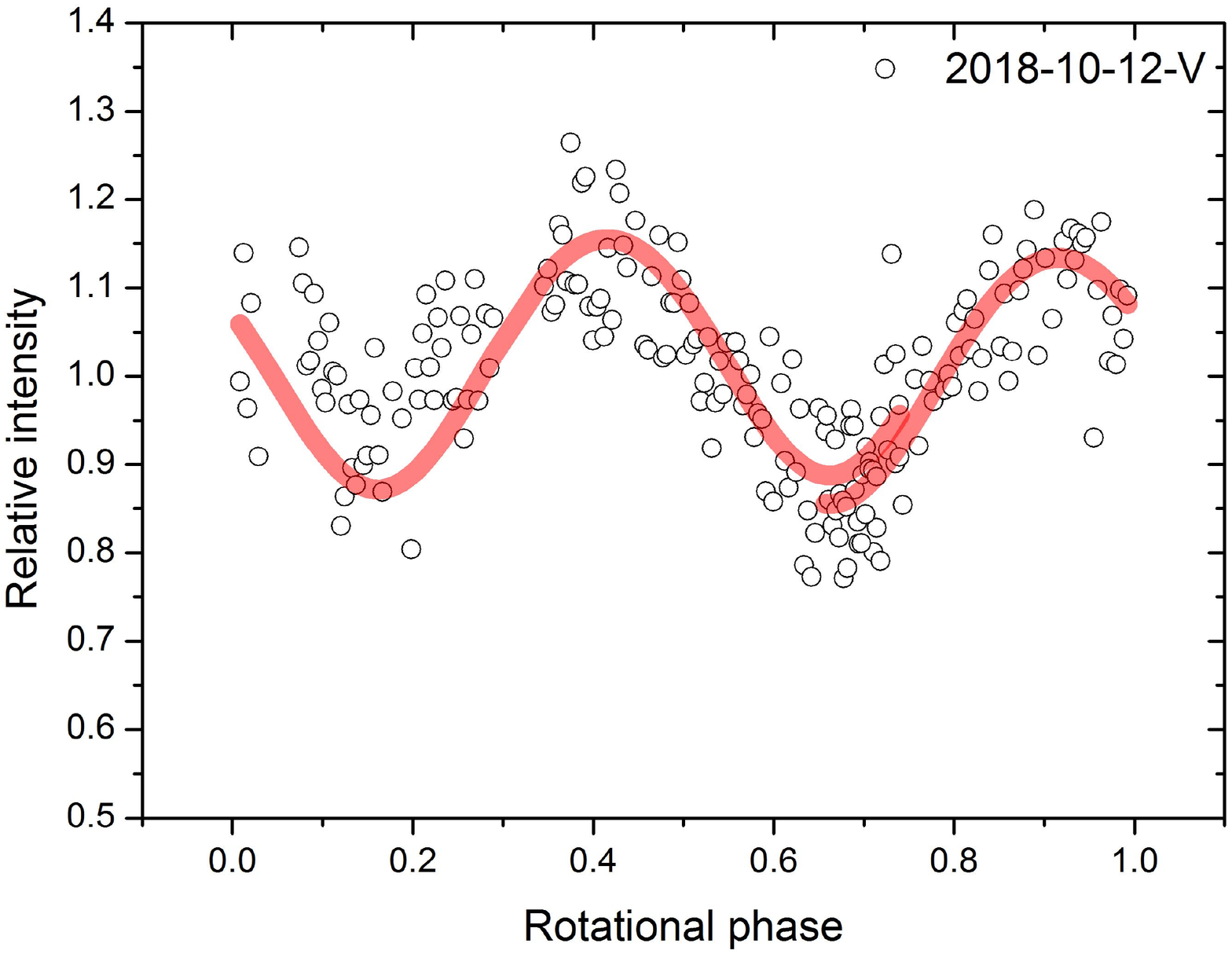}
\end{minipage}%
}%

\subfigure{
\begin{minipage}[t]{0.35\linewidth}
\centering
\includegraphics[width=5cm]{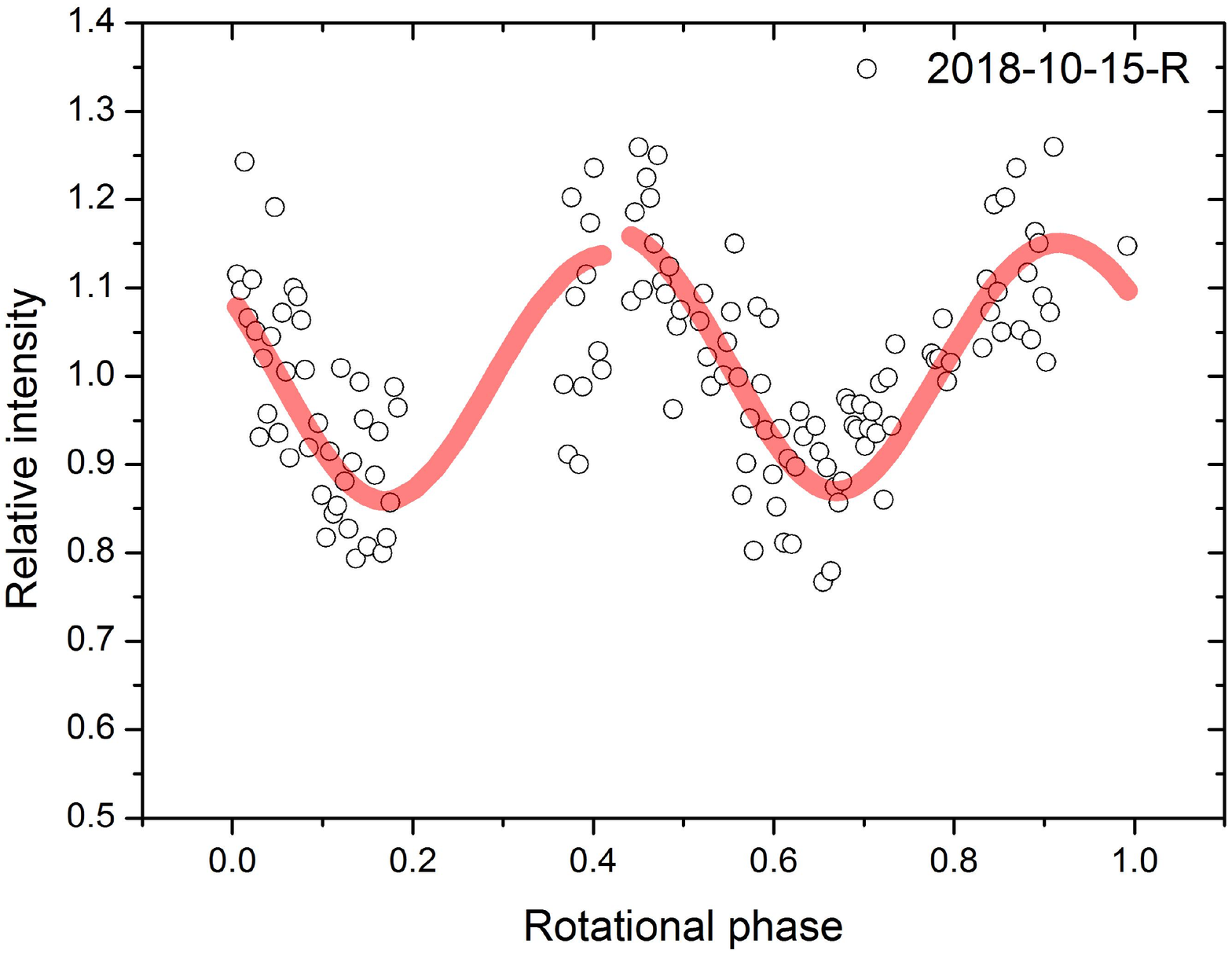}
\end{minipage}%
}%
\subfigure{
\begin{minipage}[t]{0.35\linewidth}
\centering
\includegraphics[width=5cm]{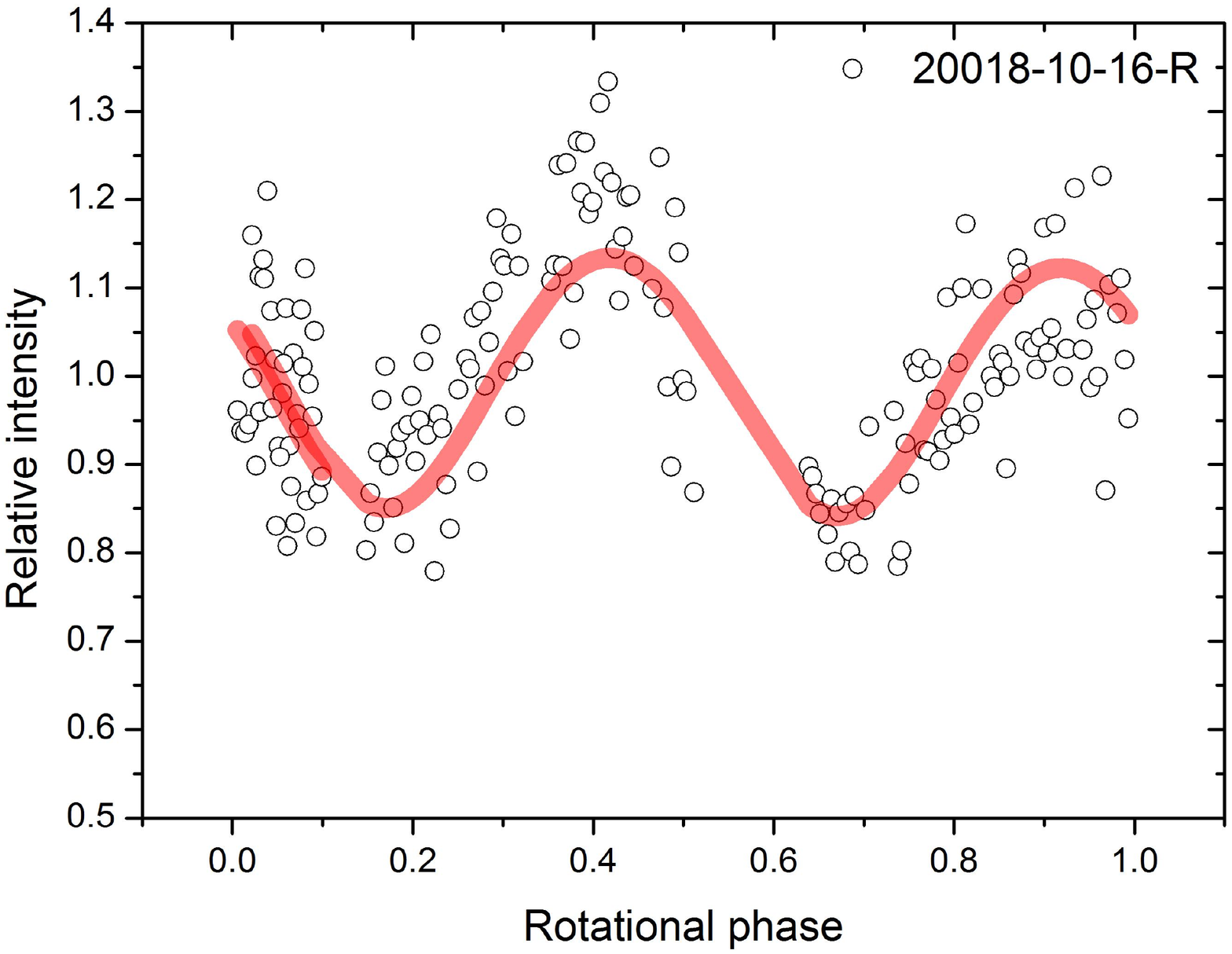}
\end{minipage}%
}%

\subfigure{
\begin{minipage}[t]{0.35\linewidth}
\centering
\includegraphics[width=5cm]{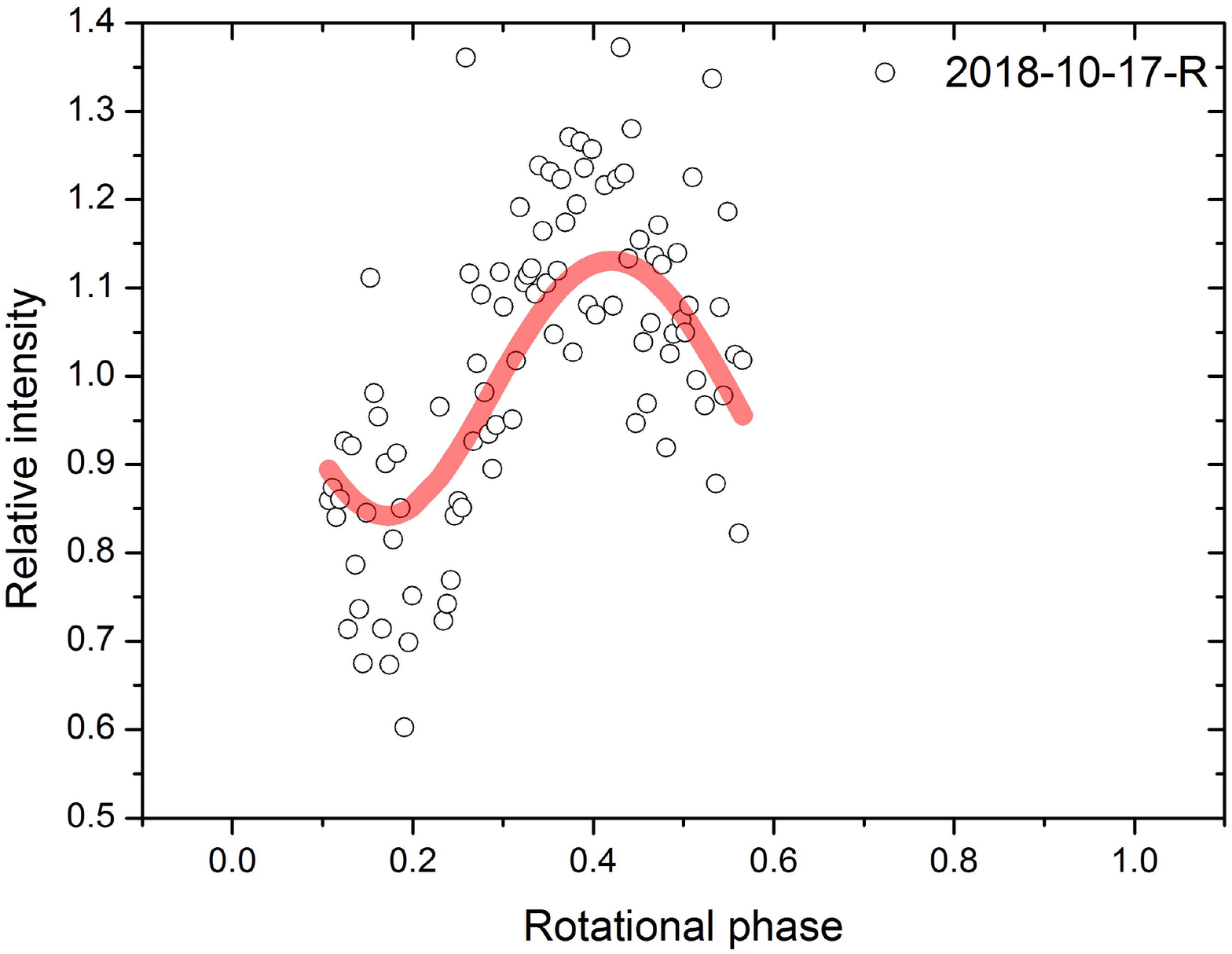}
\end{minipage}%
}%
\subfigure{
\begin{minipage}[t]{0.35\linewidth}
\centering
\includegraphics[width=5cm]{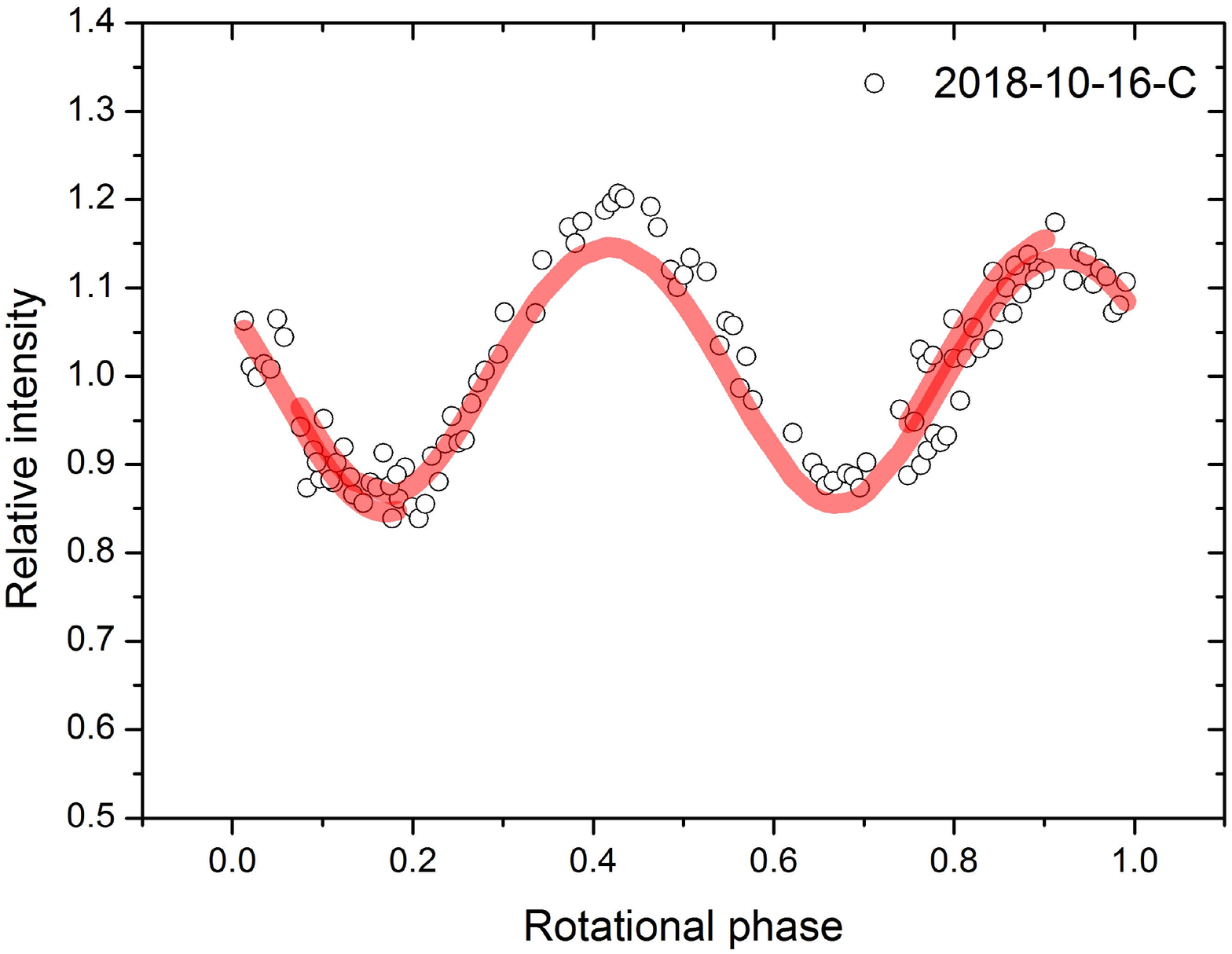}
\end{minipage}%
}%
\centering
\caption{Ten lightcurves of 2005~UD folded with the period of
  5.2340~h. Red lines are best-fit model values.}
\label{Fig1}
\end{figure}

For the phase curve analysis of 2005~UD, photometric data obtained in
different nights and/or with different filters need to be converted
into the same photometric system, e.g., the standard V-band
magnitudes. In this work, we firstly transformed the instrumental
magnitudes of celestial objects into the $r'$ band of the Carlsberg
Meridian Catalogue 15 (CMC15). The relationship between the
instrumental magnitude and the $r'$-band of CMC15 (Eq.~2) is derived
by using stable reference stars in the images. The reference stars are
chosen based on a threshold in the intrinsic variance, e.g., 0.01~mag
in the case of 2005~UD's observations. The intrinsic variance of the
stars $\sigma_{s(i)}$ in a night and the variance of the observations
$\sigma_{t(j)}$ (the variance of reference stars' magnitudes in a
image after removing their means), are output by the coarse
de-correlation method (for details, see
\cite{2006MNRAS.373..799C}). Briefly, the two variances above are
estimated iteratively by minimizing the value of $\chi^2$ (Eq.~1a
below), providing zero points of the stars' magnitudes $\hat{M}_{i}$
in a night and the zero point of each frame $\hat{Z}_{j}$ (calculated
with Eqs.~1b and 1c). The quantities $\hat{M}_{i}$ and $\hat{Z}_{j}$
are re-computed when we have new values for $\sigma_{s(i)}$ and
$\sigma_{t(j)}$. The iterative procedure is ended when the four
quantities above no longer change significantly. For a given reference
star $i$, the errors in the reduced magnitudes $r_{i,j}$ in a night,
$j$ denoting the index of a frame, are finally random, essentially
realizations of white noise. The following equation gives a
mathematical description of the procedures outlined above:
\begin{equation}
\begin{array}{lcr}
\chi^{2}=\sum\frac{(m_{ij}-\hat{M}_{i}-\hat{Z}_{j})^{2}}{\sigma^{2}_{ij}+\sigma^{2}_{s(j)}+\sigma^{2}_{t(i)}}, & & (a) \\ \\\hat{M}_{i}=\frac{\sum_{i}(m_{ij}-\hat{Z}_{j})*w_{ij}}{w_{ij}},&w_{ij}=\frac{1}{\sigma^{2}_{ij}+\sigma^{2}_{t(j)}},&(b)\\
\hat{Z}_{j}=\frac{\sum_{i}(m_{ij}-\hat{M}_{i})*u_{ij}}{u_{ij}},&u_{ij}=\frac{1}{\sigma^{2}_{ij}+\sigma^{2}_{s(i)}},&(c)\\
r_{ij}=m_{ij}-\hat{M}_{j}-\hat{Z}_{i}.&&(d)
\end{array}
\end{equation}
Here $m_{ij}$ represents the observed magnitude of star $i$ in frame
$j$, and $\sigma_{ij}$ is the corresponding observational uncertainty.

The magnitude zero points of selected reference stars in a night
$\hat{M}_{i}$ are applied to fit the relationship of Eq.~2:
 \begin{equation}
  \hat{M}_{i}{(C/R/V)}=M_{(C/R/V)0}+M_{(r')i}+ k_{(C/R/V)}\times(J_i -Ks_i)
  \end{equation}
  In detail, the parameters $M_{(C/R/V)0}$ and the color index
  coefficient $k_{(C/R/V)}$ in Eq.~2 are fitted by comparing
  the $\hat{M}_{i}$ values of selected stars to their values $M_{(r')}i$ in
  CMC15. The color indices $(J_i-Ks_i)$ of reference stars come from
  the 2MASS catalogue. The $r'$-band of CMC15 is the same as that in
  the Sloan Digital Sky Survey.  In Eq.~2, $i$ denotes the reference
  star and C, R, and V denote different filters.
  \renewcommand{\thefootnote}{\fnsymbol{footnote}}
Using the derived
  parameters $M_{(C/R/V)0}$ and $k_{(C/R/V)}$, the mean magnitude of
  the asteroid in a given night is transformed into the $r'$-band of
  the CMC15 photometric system, denoted by $M(r')_{ast}$. For no
  measurement value for 2005~UD's color index $(J-Ks)_{ast}$ , we
  temporally use that of Phaethon's at 0.275
  \footnote{https://sbnapps.psi.edu/ferret/SimpleSearch/results.action}. Then
  the $r'$-band magnitudes of the asteroid are converted into the
  standard V-band magnitude by another linear relationship
  (\cite{2009JBAA..119..149D})
 \begin{equation}
   V_{ast} = 0.6278\times (J-Ks)_{ast} + 0.9947\times M (r')_{ast}.
 \end{equation}

\begin{figure}[ht]
  \centering
  \includegraphics[width=\hsize]{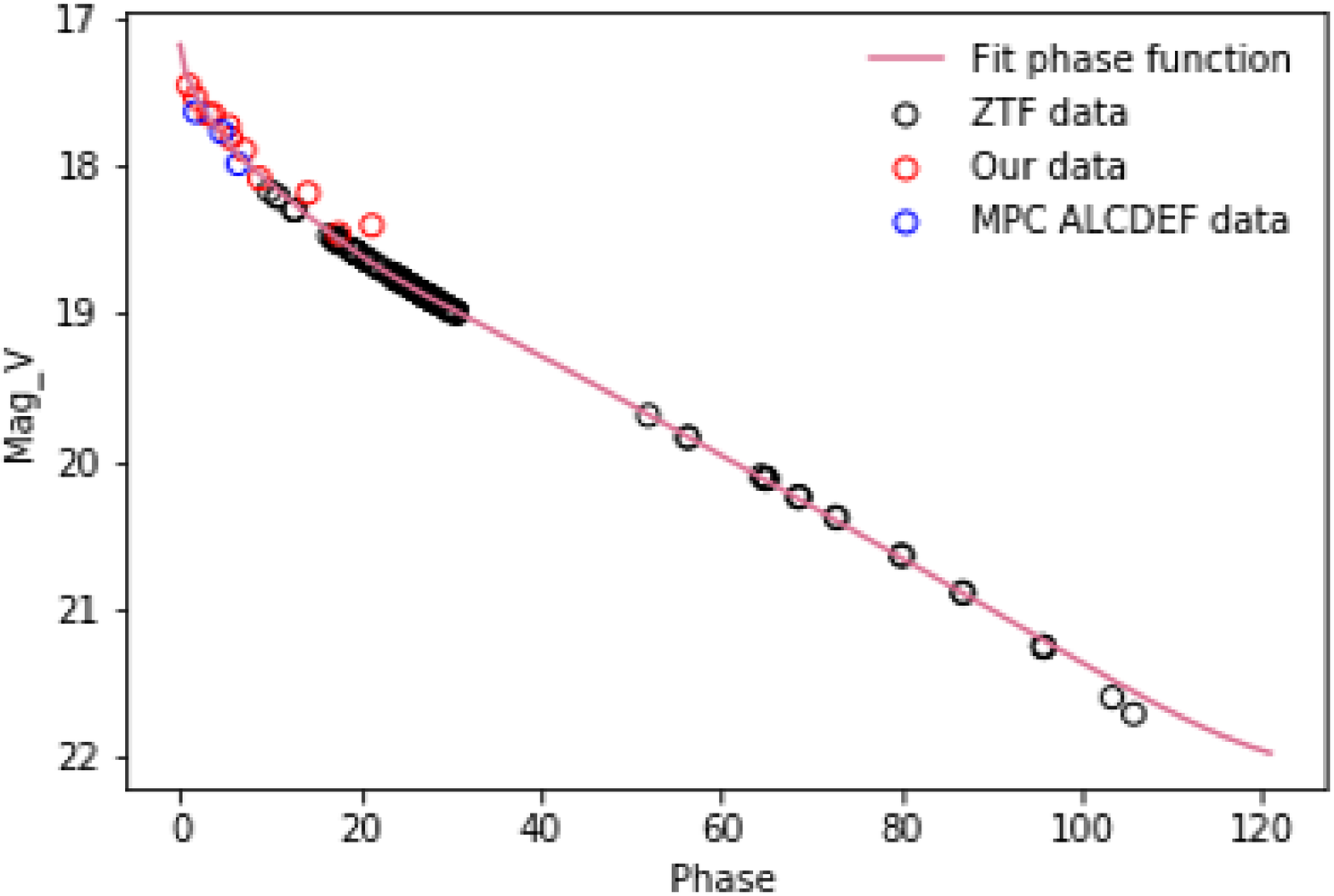}
  \caption{Phase curve of 2005~UD.}
  \label{Fig2}
\end{figure}

 The calibrated mean magnitude $V_{ast}$ of each lightcurve of 2005~UD
 are used in the phase curve inversion: the results are shown in
 Fig.~\ref{Fig2}. The individual data points of the asteroid in a night are
 calibrated by adding the calibrated mean magnitude $V_{ast}$ into the
 reduced magnitudes in that night.

 In addition, we have downloaded, from the MPC Asteroid Lightcurve
 Photometry
 Database\footnote{http://alcdef.org/PHP/alcdef$\_$GenerateALCDEFPage.php},
 three dense lightcurves of 2005~UD observed on 12, 15, and 16 October
 2018. These data have been observed through a clear filter and have
 been converted into the V-band. We corrected the distance effects on
 the magnitude and light travel time in each recorded time stamp.

 Finally, we have downloaded sparse photometric data of 2005~UD from
 the ZTF Data
 release\footnote{https://irsa.ipac.caltech.edu/applications/ztf/}. The
 ZTF is a robotic time-domain sky survey using the Samuel Oschin
 48-inch Schmidt telescope with a new Mosaic CCD of a 47 square-degree
 FOV. This observational system is equipped with three filters: ZTF-g,
 ZTF-r, and ZTF-i. The limiting magnitude in the ZTF-r band is
 20.7~mag with a $5-\sigma$ detection threshold in a 30-s exposure. It
 can scan more than 3750~deg$^{2}$ per hour. The ZTF data reduction
 follows the data processing system of the Palomar Transient Factory
 (PTF) survey. They use a fixed aperture of 8 pixels to obtain
 instrumental magnitudes and carry out photometric calibration with
 the reference stars in the SDSS catalogue and in the ZTF images
 (\cite{2012PASP..124...62O},\cite{2014PASP..126..674L},\cite{2019PASP..131a8002B}). The
 downloaded 166 photometric data points of 2005~UD span from 2017
 October to 2019 July. Those data points have been converted into the
 mean standard V band
 (\cite{2012PASP..124...62O},\cite{2019PASP..131a8002B}). For those
 data, we have corrected the distance effects on the magnitude and
 light travel time in the recorded time stamps (see Fig.~\ref{Fig2};
 the black circles denote the ZTF data).

\section{Photometric analysis}

\subsection{Lommel-Seeliger elliposid model}

An asteroid's brightness results from the scattering of sunlight by
its surface. How bright the asteroid truly is depends on its size,
shape, orientation, and surface scattering properties. The surface
scattering properties determine the asteroid's geometric albedo. The
brightness is standardized by locating, fictitiously, the asteroid at
a 1-au distance from the Sun as well as from the observer. As for the
effects on magnitude due to the varying distance of asteroid from the
Sun and the observer, we account for them by the formula
$-5\log(r\Delta)$.  The brightness of the asteroid varies with the
solar phase angle, the angle between the Sun and the observer as seen
from the object. Such brightness variation is called the phase curve.
Most asteroids have an irregular shape. For a spinning nonspherical
asteroid, the brightness varies as a function of rotational phase due
to the varying illuminated and visible parts of surface. The resulting
brightness curve is called the lightcurve.  The shape of the
lightcurve of an asteroid is usually dominated by the shape of the
asteroid. The lightcurve shape varies from one apparition to another
due to the changes in the so-called aspect angles, the angles between
the spin pole direction and the line of sight or the direction of
sunlight.  The lightcurve of an asteroid contains information on the
asteroid's physical properties, that is, size, albedo, spin status,
and shape, and it depends on the observation geometry. We may
infer the asteroid's properties from the lightcurves if the
lightcurves span a sufficiently large selection of observational
geometries.  The procedure is called lightcurve inversion, in which a
brightness model is established to invert the asteroid's properties
assuming a shape model and a surface scattering law.  Normally,
scattering laws such as the Lommel-Seeliger law, Lambert law, Hapke
model (\cite{2012Icar..221.1079H}) and Lumme-Bowell model
(\cite{1979aste.book..132B}) are used. As for the shape model, a
triaxial ellipsoid model or a more complex convex shape model can be
introduced, depending on the characteristics of the lightcurves.

The regular shape of lightcurves of 2005~UD (Fig.~\ref{Fig1}) with two
peaks implies a regular shape of the asteroid. That is the main reason
why we use, in the lightcurve inversion, the triaxial ellipsoid model
with the Lommel-Seeliger scattering law (LS ellipsoid;
\cite{2015P&SS..118..227M}). For an elementary facet $dA$ on the
asteroid's surface, the brightness of the facet with the
Lommel-Seeliger law can be written as
\begin{equation}
dL= \frac{1}{4}F_0\mu_0\varpi_0P(\alpha)\frac{1}{\mu+\mu_0}\mu dA,
\end{equation}
where $\pi F_0$ is the incident solar flux density on the facet,
$\varpi_0$ is the single-scattering albedo, $\alpha$ is the solar
phase angle, and $P(\alpha)$ is the single-scattering phase function.

The brightness integrated over an ellipsoid surface with the
Lommel-Seeliger scattering law is given as
{\setlength\arraycolsep{2pt}
 \begin{eqnarray}\label{eq5}
L(\alpha,e_\odot,e_\oplus)&=&\frac{1}{8}\pi F_0 abc \varpi_0P(\alpha)\frac{S_\odot S_\oplus}{S}\{\cos(\lambda'-\alpha')+\cos\lambda'+\sin\lambda'\sin(\lambda'-\alpha'){}
                                            \nonumber\\
 &&{}\times \ln[\cot\frac{1}{2}\lambda'\cot\frac{1}{2}(\alpha'-\lambda')]\}.
 \end{eqnarray}
Here, $e_\odot$ and $e_\oplus$ denote the unit vectors of solar and
 viewer directions, $a$, $b$, and $c$ are the three semimajor axes of
 the ellipsoid, and the auxiliary quantities $S_\odot$, $S_\oplus$,
 $\alpha'$, $\lambda'$ are functions of the illumination and viewing
 geometry, pole orientation, and shape of the asteroid. For details,
 the reader is refered to Eqs. 11 and 12 in \cite{2015P&SS..118..227M}.

Using the relationship
 \begin{displaymath}
 \frac{1}{8}\varpi_0 P(\alpha)=p\frac{\phi_{HG_1G_2}(\alpha)}{\phi_{LS}(\alpha)},
 \end{displaymath}
Eq.~\ref{eq5} can be re-written as
{\setlength\arraycolsep{2pt}
\begin{eqnarray}\label{eq6}
L(\alpha,e_\odot,e_\oplus)&=&\pi F_0 abcp\frac{\phi_{HG_1G_2}(\alpha)}{\phi_{LS}(\alpha)}\frac{S_\odot S_\oplus}{S}\{\cos(\lambda'-\alpha')+\cos\lambda'+\sin\lambda'\sin(\lambda'-\alpha'){}
    \nonumber\\
&&\times \ln[\cot\frac{1}{2}\lambda'\cot\frac{1}{2}(\alpha'-\lambda')]\},
\end{eqnarray}
where
\begin{displaymath}
\phi_{LS}(\alpha) = 1- \sin\frac{1}{2}\alpha \tan\frac{1}{2}\alpha \ln(\cot\frac{1}{4}\alpha).
\end{displaymath}
$\phi_{LS}(\alpha)$ is the Lommel-Seeliger disk-integrated phase
function for a spherical asteroid and $p$ is the geometric albedo. In
the model above, the $H,G_1,G_2$ phase function
(\cite{2010Icar..209..542M}) is incorporated into the LS-ellipsoid
brightness model.

In short, altogether 12 unknown parameters are involved in the
LS-ellipsoid model. They are the rotational period $Per$, pole
orientation ($\lambda, \beta$) in the ecliptic frame of J2000.0,
rotational phase $\varphi_0$ at $JD_0$, three semimajor axes
$(a, b, c)$, geometric albedo $p$, phase function parameters
$H,G_1, G_2$, and the equivalent diameter of asteroid $D$. To derive
the solution of those unknown parameters, the flexible Nelder-Mead
downhill method and a Markov-chain Monte Carlo method (MCMC) are
applied in our photometric analysis procedure. In practice, the
analysis procedure of 2005~UD consists of two parts: the shape
inversion of 2005~UD with the LS-ellipsoid model and the phase curve
inversion. In the first part, we estimate rotation period, pole
longitude and latitude, and three semimajor axes using 14 dense
lightcurves. In the second part, using calibrated photometric data of
2005~UD, the phase curve parameters $H, G_1, G_2$ are retrieved.

\subsection{Shape inversion}

Altogether 14 dense lightcurves of 2005~UD are used to invert the spin
and shape parameters by the Nelder-Mead downhill simplex
method. Furthermore, the uncertainties of the parameters are assessed
by the MCMC method.

Using the downhill simplex least-squares method, the following
$\chi^{2}(Par)$ is minimized:
\begin{equation}
\chi^{2}(Par)=\sum_{i=1}^{N_0}\sum_{j=1}^{N_i}\frac{1}{\sigma_{ij}^{2}}[L_{obs,ij}-L_{ij}(Par)]^{2}.
\end{equation}
In practical shape inversion, only 7 parameters ($Per$, $\lambda $,
$\beta $, $\varphi_0 $, $a$, $b$, $c$) are estimated and the rest of the
parameters are kept fixed in Eq.~7. Here $N_0$ is the number of
lightcurves used, $N_i$ is the number of data points in the $i$th
lightcurve, $L_{obs,ij}$ is the $j$th data point in the $i$th
lightcurve, and $\sigma_{ij}$ is its corresponding
uncertainty. $L_{ij}(Par)$ is the modeled brightness calculated with
the LS-ellipsoid model.

To find the most probable rotation period, a wide range of periods
within $2.5-12.5$~h was scanned with a step of $Per^{2}/2T$ (here
$Per$ is the assumed period and $T$ is the time span of all involved
data). Fig.~\ref{Fig3} shows the $\chi^2$ of
the lightcurve fitting versus the trial period, the most likely value of period is located at
5.2338~h. The result is close to Kinoshita's result of 5.2310~h
(\cite{2007A&A...466.1153K}).  First, at each step of period scanning,
hundreds of different initial poles distributed uniformly over the
unit sphere are tested. Second, we have scanned the entire 
unit sphere with a step of $1^\circ$ in longitude and latitude
directions to find the most probably pole of 2005~UD using the period
of 5.2338~h as the initial value. The contours of $\chi^2$ versus the
trial poles are shown in Fig.~\ref{Fig4}. The areas in blue color in
Fig.~\ref{Fig4} are corresponding to \emph{relatively small} $\chi^2$. Two
candidate poles of $(73^\circ,-84^\circ)$ and $(285^\circ,-21^\circ)$
are found with almost equal values of $\chi^2$. Third, taking the
scanned spin parameters as initial values, unknown parameters are
resolved with the Nelder-Mead downhill simplex method. Finally, we
arrive at the following pair of pole solutions: Pole 1 at
($72^\circ.6^, -84^\circ.6$) with axial ratios of $b/a=0.76$,
$c/a=0.40$ and Pole 2 at ($285^\circ.8,-25^\circ.8$) with axial ratios
of $b/a=0.76$ and $c/a=0.40$. The periods corresponding to the poles
are close to 5.2338~h.

\begin{figure}[ht]
  \centering
  \includegraphics[width=\hsize]{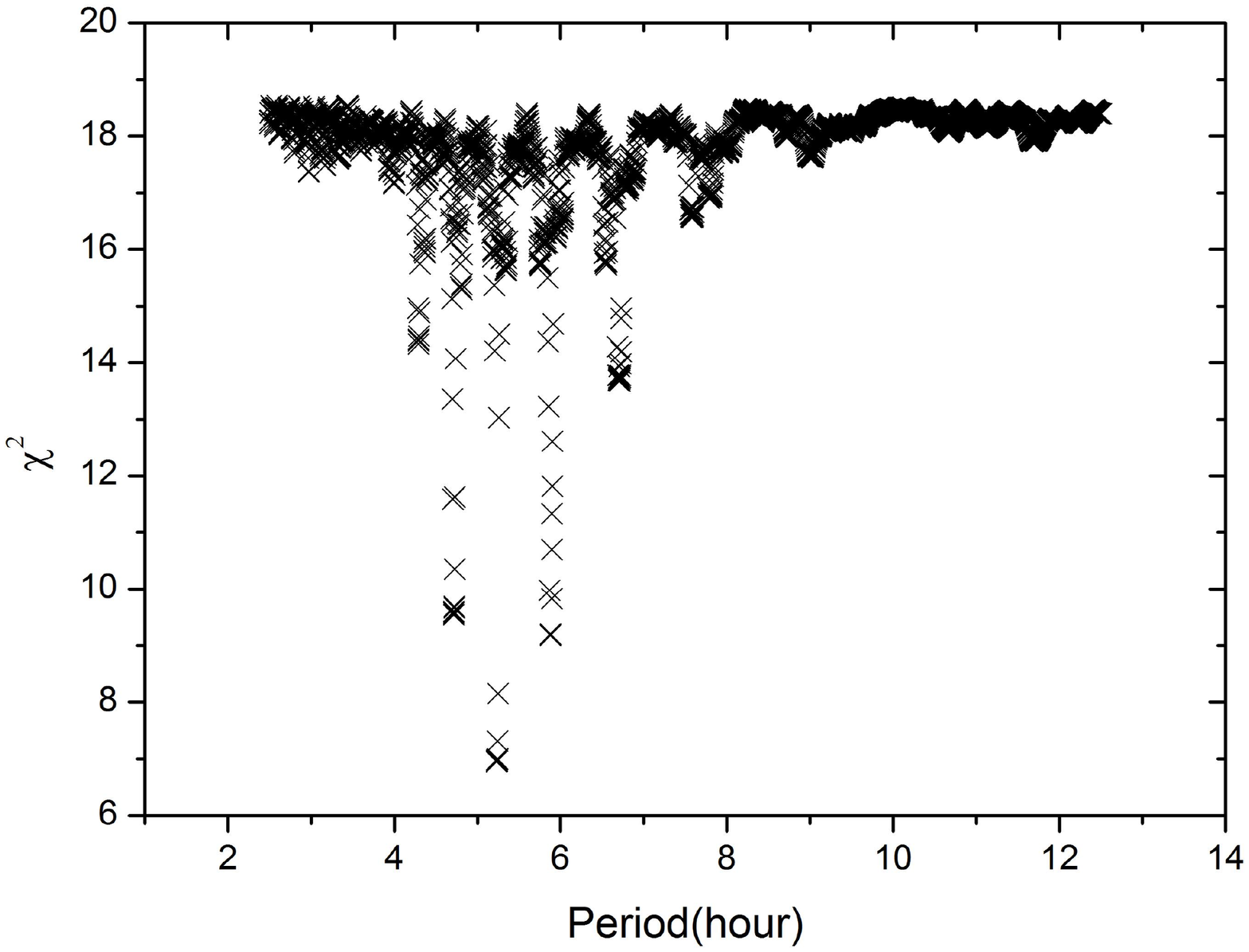}
  \caption{The $\chi^{2}$-values for the trial periods.}
  \label{Fig3}
\end{figure}

\begin{figure}[ht]
  \centering
  \includegraphics[width=\hsize]{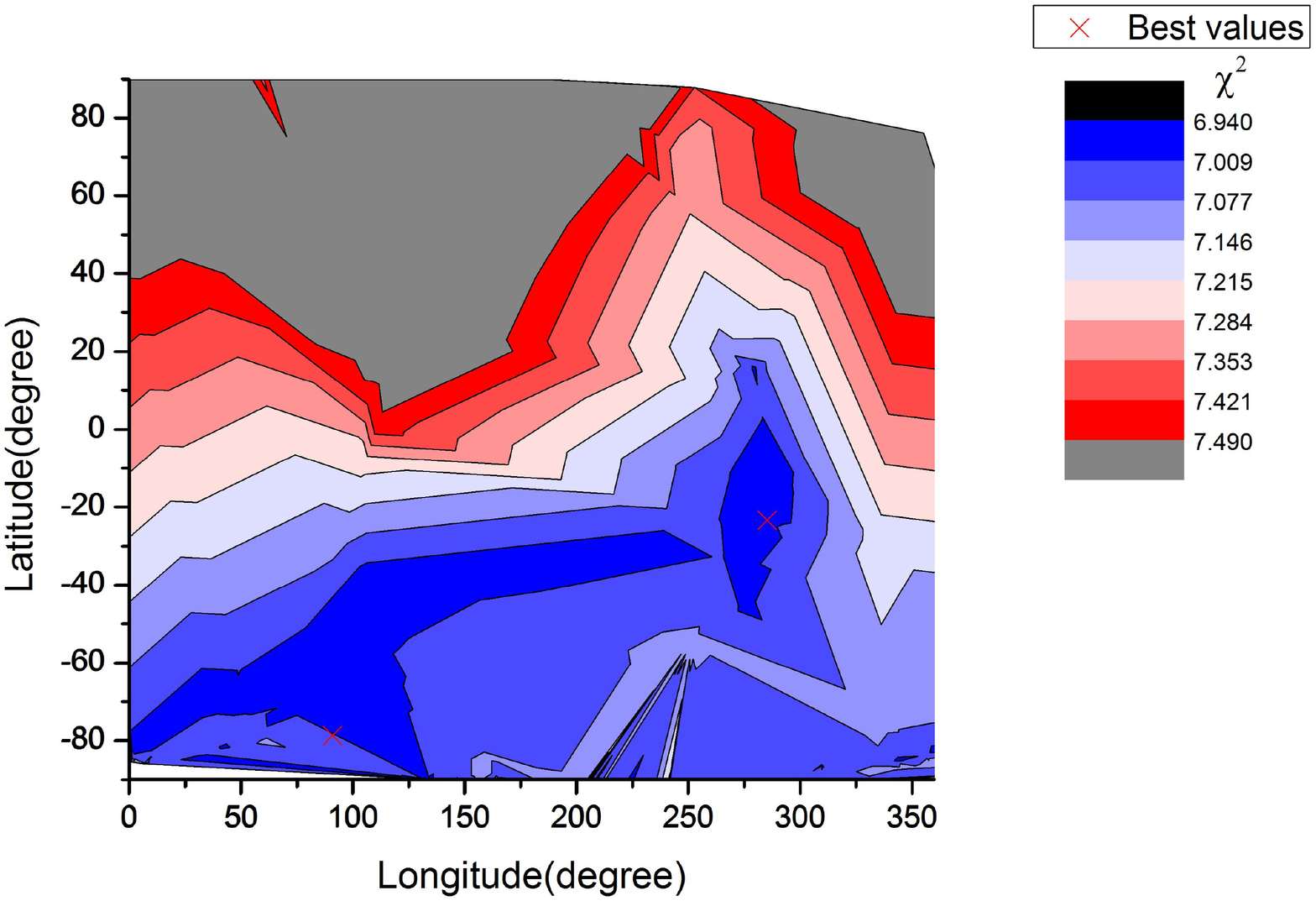}
  \caption{The $\chi^{2}$-values for  the trial pole orientations.}
  \label{Fig4}
\end{figure}

In order to derive the uncertainties for the spin and shape parameters
of 2005~UD,an MCMC simulation is run based on the photometric data of
2005~UD and the LS-ellipsoid model. The a posteriori probability
density for the parameters is characterized by a large number of
sample solutions obtained by Metropolis-Hastings sampling. The
proposal densities for the parameters are constructed via a collection
of virtual least-squares solutions derived from virtual photometric
data.} The virtual photometric data are generated by adding Gaussian
noise into the observations. At least 10000 samples are obtained with
the MCMC simulation. The joint distributions of the spin parameters
are shown in Fig.~\ref{Fig8} for Pole 1 and Fig.~\ref{Fig9} for Pole
2. The dotted lines in Figs.~\ref{Fig8} and \ref{Fig9} are the
best-fit values of the parameters. The intervals between the best-fit
value and the $1-\sigma$ limits for each distribution are used to
estimate the uncertainties of the parameters. Based on those joint
distributions, the best-fit values of the parameters with their
uncertainties are as follows: Pole 1 at
($ 72^\circ.6^{+4.2}_{-7.3}, -84^\circ.6^{+6.2}_{-2.1}$) with axial
ratios $b/a=0.75^{+0.01}_{-0.01}$ and $c/a=0.40^{ +0.16}_{-0.01}$ and
Pole 2 at ($285^\circ.8^{+1.1}_{-5.3}$, $-25^\circ.8^{+5.3}_{-12.5}$)
with axial ratios $b/a=0.76^{+0.01}_{-0.01}$ and
$c/a=0.40^{ +0.03}_{-0.01}$,}. The periods corresponding to the poles
are $5.23351^{+0.00002}_{-0.00001}$ and
$5.23403^{+0.00004}_{-0.00001}$, respectively. Comparing the
distributions of the poles, the solution of Pole 2 is
prefered. Indeed, which one of the pole solutions is the true solution
requires more observations, even observations with other techniques
(e.g., imaging, occultation, or radar).

In order to understand the inversion results intuitively, the model
brightnesses from the solution with Pole 2 solution are shown in
Fig.~\ref{Fig1} together with the observations. Most observed
brightnesses are fitted well by the modeled brightness, but some data
(e.g., data obtained on on 10 Oct. 2018) are not. This minor caveat
may be due to the fast sky-plane motion of the asteroid, resulting in
a low quality of photometric data due to the elongated image of the
asteroid.

\begin{figure}[ht]
  \centering
  \includegraphics[width=19cm]{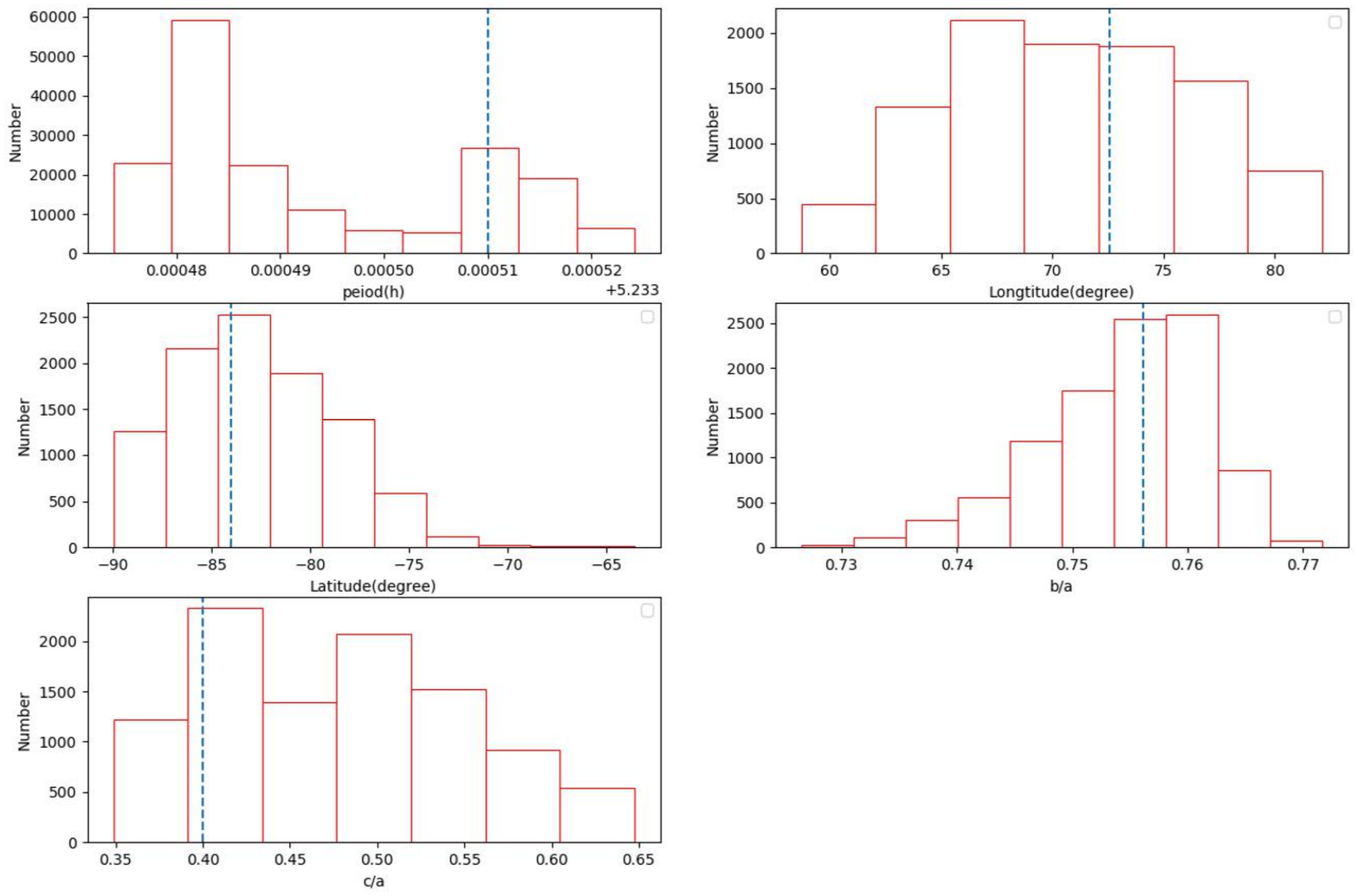}
  \caption{Joint distributions of spin and shape parameters for Pole 1.}
  \label{Fig8}
\end{figure}

\begin{figure}[ht]
  \centering
  \includegraphics[width=19cm]{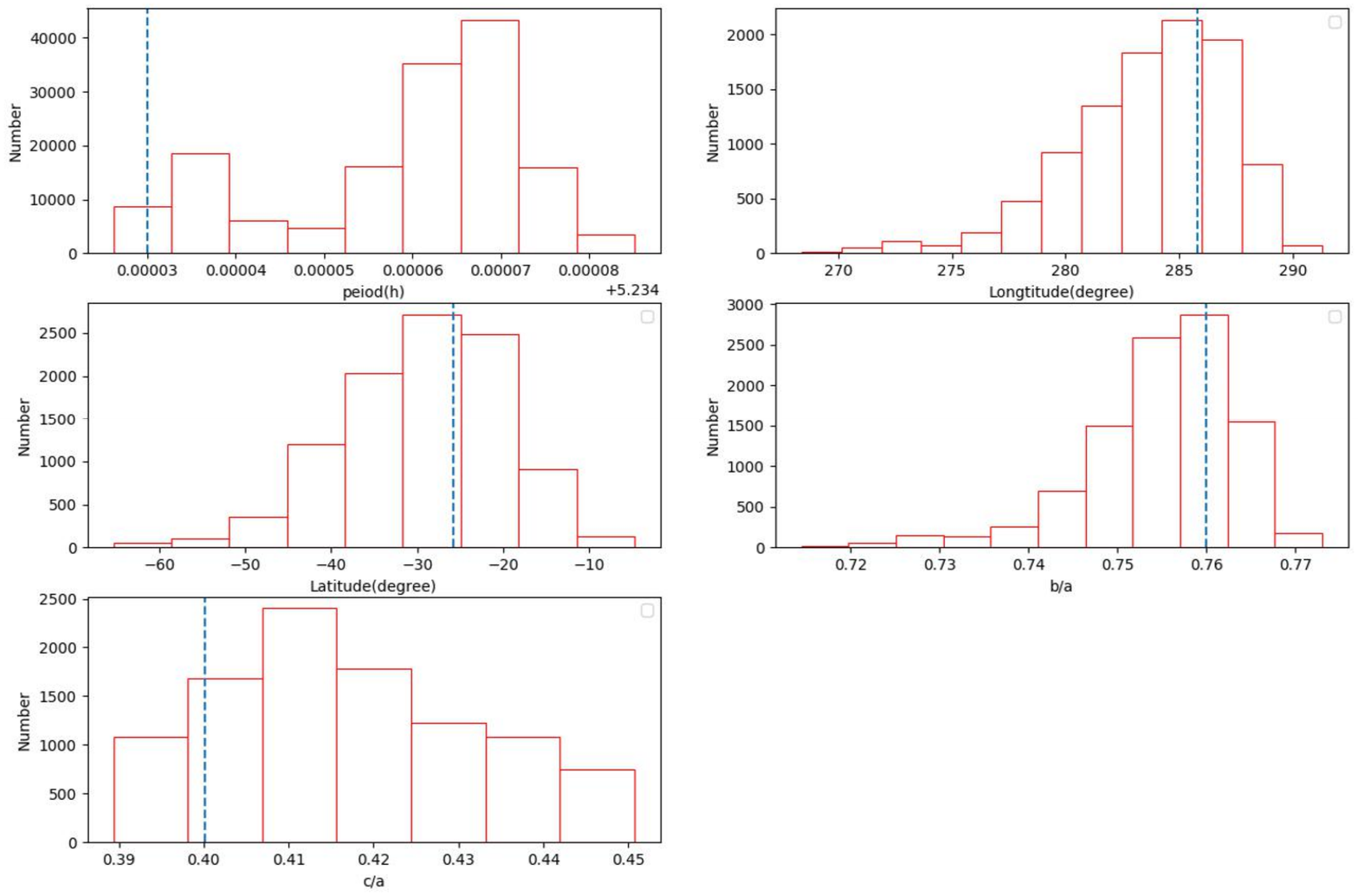}
  \caption{Joint distributions of spin and shape parameters for Pole 2.}
  \label{Fig9}
\end{figure}

\subsection{ Phase curve fitting}

The photometric phase curve of an asteroid shows the observed
brightness variation as a function of the solar phase angle. There are
two phase functions, the $H,G$ and $H,G_1,G_2$ phase functions, to be
used to describe the photometric phase curves of asteroids. The $H,G$
system ($H$ and $G$ being the absolute magnitude and slope parameter
of the asteroid, respectively) has been adopted by the International
Astronomical Union (IAU) in 1985 as the standard photometric system
for asteroids, developed from the Lumme-Bowell model
(\cite{1989aste.conf..524B}).  The slope parameter $G$ characterizes
an asteroid's surface, whereas the absolute magnitude $H$ is related
to its size and geometric albedo. The $H,G_1,G_2$ phase function is
used in the standard photometric system of asteroids adopted by the
IAU in 2012. The new three-parameter phase function improves the fits
of phase curves of both high-albedo and low-albedo asteroids. Here, we
use the $H,G_1,G_2$ phase function to fit the calibrated photometric
data of 2005~UD. The basic model for linear least-squares fitting is
{\setlength\arraycolsep{2pt}}
\begin{eqnarray}\label{eq6}
  10^{-0.4V(\alpha)}&=&a_1\phi_1(\alpha)+a_2\phi_2(\alpha)+a_3\phi_3(\alpha){}
                        \nonumber \\
                    &=&10^{-0.4H}[G_1\phi_1(\alpha)+G_2\phi_2(\alpha)+(1-G_1-G_2)\phi_3(\alpha)).
 \end{eqnarray}
 The source code for $H,G_1,G_2$ phase curve inversion is openly
 available \footnote{http://h152.it.helsinki.fi/HG1G2/}. In practice,
 the phase curve inversion is carried out based on the disk-integrated
 brightnesses of the asteroid (see Eq.~\ref{eq6}). The linear
 parameters $a_1,a_2$, and $a_3$ are derived first, then the
 parameters $H, G_1$, and $G_2$ are calculated using Eq. 19 in
 \cite{2010Icar..209..542M}. The functions $\phi_{1,2,3}(\alpha)$ are
 basis functions expressed in cubic splines with the interpolation
 grid of phase angles of
 $(0^{\circ}, 0.3^{\circ}, 1^{\circ}, 2^{\circ}, 4^{\circ}, 8^{\circ},
 15^{\circ}, 30^{\circ}, 60^{\circ}, 90^{\circ}, 120^{\circ}$, and
 $150^{\circ})$. The detailed values are available in Tables 3 and 4
 in \cite{2010Icar..209..542M}).

\begin{figure}[ht]
  \centering
  \includegraphics[width=19cm]{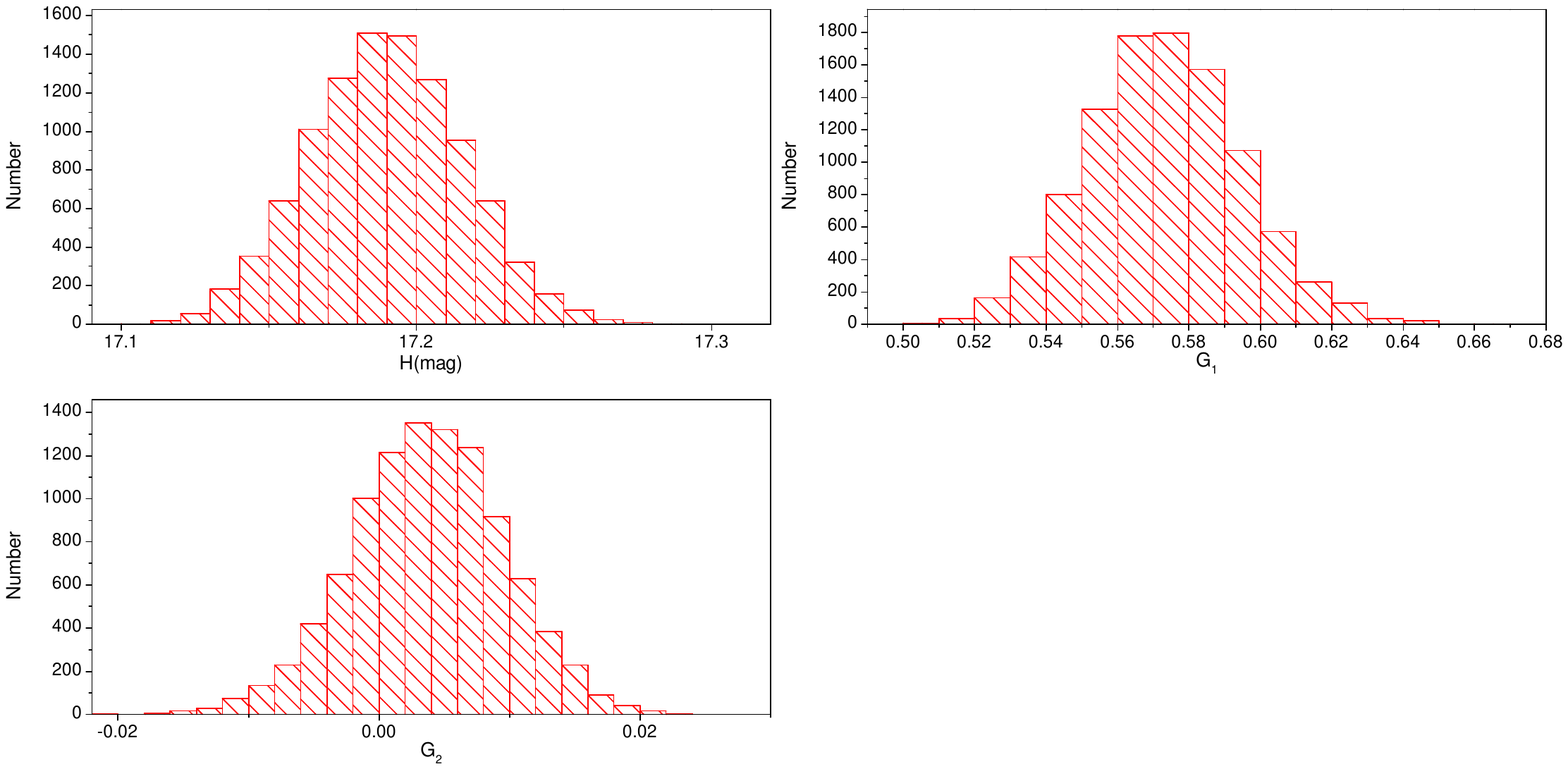}
  \caption{Marginal distributions of $H, G_1,G_2 $ for 2005 UD.}
  \label{Fig7}
\end{figure}

 The mean values of the dense lightcurves and ZTF data are used to fit
 the $H, G_1,G_2$ function with the MCMC method. In the MCMC
 simulation procedure, the Metropolis-Hastings algorithm is applied to
 sample using the Gaussian proposal probability densities of the
 parameters. Fig.~\ref{Fig7} shows the joint distributions of the
 three parameters. The best values of $H, G_1$, and $G_2$ are
 17.19~mag, 0.573, and 0.004. Using $1-\sigma$ limits of the
 distributions, we obtain $H = 17.19^{+0.10}_{-0.09}$~mag,
 $G_1=0.573 ^{+0.088}_{-0.069}, G_2=0.004^{+0.020}_{-0.021}$. The
 best-fit model to the photometric data is displayed in
 Fig.~\ref{Fig2}. From the best-fit values of $G_1$ and $G_2$, 2005~UD
 is likely to be a C-type asteroid according to the suggestion in
 \cite{2016P&SS..123..101S}.

 \section{Summary }

 In order to study the near-Earth asteroid 2005 UD, we carried out 11
 nights of photometric observations with a 30-cm telescope at the
 Corona borealis Observatory in Ali, Tibet, China. Combining our 11
 lightcurves with the 3 lightcurves from the MPC ALCDEP database, the
 spin and shape parameters of 2005~UD have been analysed with the
 Lommel-Seeliger ellipsoid method. Two pole solutions, Pole 1 at
 ($ 72^\circ.6^{+4.2}_{-7.3}, -84^\circ.6^{+6.2}_{-2.1}$) and Pole 2
 at ($285^\circ.8^{+1.1}_{-5.3}, -25^\circ.8^{+5.3}_{-12.5}$), have
 been derived. Comparing the distributions of spin parameters, we
 prefer Pole 2, since it gives more concentrated distributions. The
 axial ratios of the ellipsoid corresponding to Pole 2 are
 $b/a = 0.76^{+0.01}_{-0.01}, c/a = 0.40 ^{+0.03}_{-0.01}$. The spin
 period is $5.23403^{+0.00004}_{-0.00001}$ h.

 The distribution of the pole latitude (see Figs.~\ref{Fig8} and
 \ref{Fig9}) is wider than that of the pole longitude. This is due to
 a small span of the aspect angles of the photometric data. So more
 photometric observations are necessary for improving the pole
 orientation, especial in latitude.

 Our group also focuses on the physical studies of another PGC member,
 that is, asteroid (3200)~Phaethon. We have found that Pole 2 of
 2005~UD is close to that of Phaethon. Investigating published pole
 solutions of Phaethon, we have noted that they show slight
 differences: for example, ($97^{o}\pm10^{o}$,$-11^{o}\pm10^{o}$) and
 ($276^{o}\pm10^{o}$,$-15^{o}\pm10^{o}$) from
 \cite{2002Icar..158..294K}, $(85^{o}, -20^{o})$ from
 \cite{2014ApJ...793...50A}, $(319^{o}, -39^{o})$ from
 \cite{2016A&A...586A.108H}, and
 ($308^{o}\pm10^{o}$,$-52^{o}\pm10^{o}$) and ($322^{o}\pm10^{o}$,
 $-40^{o}\pm10^{o}$) from\cite{2018A&A...619A.123K}.  Our group gives
 a pair of poles of ($95^o.9, -20^o.4)$) and $(311^{o}.2,-23^{o}.6)$
 (a paper is being prepared). If considering the second pole solution
 of Phaethon, above pole solutions are around three directions:
 $(a)$($276^{o}\pm10^{o}$, $-15^{o}\pm10^{o}$),
 $(b)$$(311^{o}.2,-23^{o}.6)$ and $(c)$$(319^{o}, -39^{o})$. The
 differences among the three orientations occur in ecliptic latitude:
 $-15^o, -23^o$, and $-39^o$. In the case of (b) and (c), the
 longitudes are very close to each other, whereas in the case of (a),
 the longitude diverges from that of (b) and (c). Considering the
 uncertainty of the pole solutions, Phaethon's pole appears to align
 with that of 2005~UD. If it is true, we think this evidence implies
 2005~UD probably originated via a collision or rotational fission
 from Phaethon's parent body, resembling the case of the Koronis
 family (\cite{2002Natur.419...49S}).

 Combining the mean magnitudes of the dense lightcurves and the ZTF
 data, we have fitted the photometric phase curve of 2005~UD with the
 three parameter $H,G_1, G_2$ phase function. The parameters $H, G_1$
 and $G_2$ are as follows: $17.19^{+0.10}_{-0.09}$~mag,
 $0.573 ^{+0.088}_{-0.069}, 0.004^{+0.020}_{-0.021}$,
 respectively. Given these values of the parameters, the phase
 integral parameter $q$, normalized slope of the phase-curve $k$
 (within $\alpha( 7^{o}.5$) and the amplitude of the opposition effect
 $\zeta-1$ ($\zeta$ is the enhancement factor) are estimated to be
 0.2441, -1.9076, and 0.7309 as computed from the following
 relationships(\cite{2010Icar..209..542M}):
\begin{displaymath}
q = 0.009082 + 0.4061G_1 + 0.8092G_2,
\end{displaymath}
\begin{displaymath}
k = -\frac{G_1\frac{6}{\pi}+G_2\frac{9}{5\pi}}{G_1+G_2} = -\frac{1}{5\pi}\frac{30G_1+9G_2}{G_1+G_2},
\end{displaymath}
\begin{displaymath}
\zeta -1 = \frac{1 - G_1 -G_2}{G_1 +G_2 }.
\end{displaymath}


Based on the derived $H$ value and the relationship of the diameter
and albedo ($D = \frac{1329}{\sqrt{p_v}}10^{-0.2H}$)
\citep{1989aste.conf..524B}, the equivalent diameter $D$ of 2005~UD is
estimated to be 1.3~km using its new derived albedo of $0.14\pm 0.09$
(\cite{2019AJ....158...97M}), which is slightly higher than the
previously estimated value of $1.2\pm0.4$~km
(\cite{2019AJ....158...97M}).

\section{Acknowledgements}

The research has been funded by the National Natural Science
Foundation of China (Grant Nos. 11073051 and 11673063) and the Academy
of Finland (Grant No. 325805).  The research has made use of the
NASA/IPAC Infrared Science Archive, which is funded by the National
Aeronautics and Space Administration and operated by the California
Institute of Technology. The work includes data from the Asteroid
Terrestrial-impact Last Alert System (ATLAS) project. ATLAS is
primarily funded to search for near-Earth asteroids through NASA
grants NN12AR55G, 80NSSC18K0284, and 80NSSC18K1575; byproducts of the
near-Earth-object search include images and catalogs from the survey
area. The ATLAS science products have been made possible through the
contributions of the University of Hawaii Institute for Astronomy, the
Queen's University Belfast, the Space Telescope Science Institute, and
the South African Astronomical Observatory.




 \bibliographystyle{elsarticle-harv.bst}
 \bibliography{reference}

\begin{thebibliography}{26}
\expandafter\ifx\csname natexlab\endcsname\relax\def\natexlab#1{#1}\fi
\providecommand{\url}[1]{\texttt{#1}}
\providecommand{\href}[2]{#2}
\providecommand{\path}[1]{#1}
\providecommand{\DOIprefix}{doi:}
\providecommand{\ArXivprefix}{arXiv:}
\providecommand{\URLprefix}{URL: }
\providecommand{\Pubmedprefix}{pmid:}
\providecommand{\doi}[1]{\href{http://dx.doi.org/#1}{\path{#1}}}
\providecommand{\Pubmed}[1]{\href{pmid:#1}{\path{#1}}}
\providecommand{\bibinfo}[2]{#2}
\ifx\xfnm\relax \def\xfnm[#1]{\unskip,\space#1}\fi
\bibitem[{{Ansdell} et~al.(2014){Ansdell}, {Meech}, {Hainaut}, {Buie},
  {Kaluna}, {Bauer} and {Dundon}}]{2014ApJ...793...50A}
\bibinfo{author}{{Ansdell}, M.}, \bibinfo{author}{{Meech}, K.J.},
  \bibinfo{author}{{Hainaut}, O.}, \bibinfo{author}{{Buie}, M.W.},
  \bibinfo{author}{{Kaluna}, H.}, \bibinfo{author}{{Bauer}, J.},
  \bibinfo{author}{{Dundon}, L.}, \bibinfo{year}{2014}.
\newblock \bibinfo{title}{{Refined Rotational Period, Pole Solution, and Shape
  Model for (3200) Phaethon}}.
\newblock \bibinfo{journal}{The Astrophysical Journal} \bibinfo{volume}{793},
  \bibinfo{pages}{50}.
\newblock \DOIprefix\doi{10.1088/0004-637X/793/1/50},
  \href{http://arxiv.org/abs/1407.7886}{{\tt arXiv:1407.7886}}.
\bibitem[{{Bellm} et~al.(2019){Bellm}, {Kulkarni}, {Graham}, {Dekany}, {Smith},
  {Riddle}, {Masci}, {Helou}, {Prince}, {Adams}, {Barbarino}, {Barlow},
  {Bauer}, {Beck}, {Belicki}, {Biswas}, {Blagorodnova}, {Bodewits}, {Bolin},
  {Brinnel}, {Brooke}, {Bue}, {Bulla}, {Burruss}, {Cenko}, {Chang}, {Connolly},
  {Coughlin}, {Cromer}, {Cunningham}, {De}, {Delacroix}, {Desai}, {Duev},
  {Eadie}, {Farnham}, {Feeney}, {Feindt}, {Flynn}, {Franckowiak}, {Frederick},
  {Fremling}, {Gal-Yam}, {Gezari}, {Giomi}, {Goldstein}, {Golkhou}, {Goobar},
  {Groom}, {Hacopians}, {Hale}, {Henning}, {Ho}, {Hover}, {Howell}, {Hung},
  {Huppenkothen}, {Imel}, {Ip}, {Ivezi{\'c}}, {Jackson}, {Jones}, {Juric},
  {Kasliwal}, {Kaspi}, {Kaye}, {Kelley}, {Kowalski}, {Kramer}, {Kupfer},
  {Landry}, {Laher}, {Lee}, {Lin}, {Lin}, {Lunnan}, {Giomi}, {Mahabal}, {Mao},
  {Miller}, {Monkewitz}, {Murphy}, {Ngeow}, {Nordin}, {Nugent}, {Ofek},
  {Patterson}, {Penprase}, {Porter}, {Rauch}, {Rebbapragada}, {Reiley},
  {Rigault}, {Rodriguez}, {van Roestel}, {Rusholme}, {van Santen}, {Schulze},
  {Shupe}, {Singer}, {Soumagnac}, {Stein}, {Surace}, {Sollerman}, {Szkody},
  {Taddia}, {Terek}, {Van Sistine}, {van Velzen}, {Vestrand}, {Walters},
  {Ward}, {Ye}, {Yu}, {Yan} and {Zolkower}}]{2019PASP..131a8002B}
\bibinfo{author}{{Bellm}, E.C.}, \bibinfo{author}{{Kulkarni}, S.R.},
  \bibinfo{author}{{Graham}, M.J.}, \bibinfo{author}{{Dekany}, R.},
  \bibinfo{author}{{Smith}, R.M.}, \bibinfo{author}{{Riddle}, R.},
  \bibinfo{author}{{Masci}, F.J.}, \bibinfo{author}{{Helou}, G.},
  \bibinfo{author}{{Prince}, T.A.}, \bibinfo{author}{{Adams}, S.M.},
  \bibinfo{author}{{Barbarino}, C.}, \bibinfo{author}{{Barlow}, T.},
  \bibinfo{author}{{Bauer}, J.}, \bibinfo{author}{{Beck}, R.},
  \bibinfo{author}{{Belicki}, J.}, \bibinfo{author}{{Biswas}, R.},
  \bibinfo{author}{{Blagorodnova}, N.}, \bibinfo{author}{{Bodewits}, D.},
  \bibinfo{author}{{Bolin}, B.}, \bibinfo{author}{{Brinnel}, V.},
  \bibinfo{author}{{Brooke}, T.}, \bibinfo{author}{{Bue}, B.},
  \bibinfo{author}{{Bulla}, M.}, \bibinfo{author}{{Burruss}, R.},
  \bibinfo{author}{{Cenko}, S.B.}, \bibinfo{author}{{Chang}, C.K.},
  \bibinfo{author}{{Connolly}, A.}, \bibinfo{author}{{Coughlin}, M.},
  \bibinfo{author}{{Cromer}, J.}, \bibinfo{author}{{Cunningham}, V.},
  \bibinfo{author}{{De}, K.}, \bibinfo{author}{{Delacroix}, A.},
  \bibinfo{author}{{Desai}, V.}, \bibinfo{author}{{Duev}, D.A.},
  \bibinfo{author}{{Eadie}, G.}, \bibinfo{author}{{Farnham}, T.L.},
  \bibinfo{author}{{Feeney}, M.}, \bibinfo{author}{{Feindt}, U.},
  \bibinfo{author}{{Flynn}, D.}, \bibinfo{author}{{Franckowiak}, A.},
  \bibinfo{author}{{Frederick}, S.}, \bibinfo{author}{{Fremling}, C.},
  \bibinfo{author}{{Gal-Yam}, A.}, \bibinfo{author}{{Gezari}, S.},
  \bibinfo{author}{{Giomi}, M.}, \bibinfo{author}{{Goldstein}, D.A.},
  \bibinfo{author}{{Golkhou}, V.Z.}, \bibinfo{author}{{Goobar}, A.},
  \bibinfo{author}{{Groom}, S.}, \bibinfo{author}{{Hacopians}, E.},
  \bibinfo{author}{{Hale}, D.}, \bibinfo{author}{{Henning}, J.},
  \bibinfo{author}{{Ho}, A.Y.Q.}, \bibinfo{author}{{Hover}, D.},
  \bibinfo{author}{{Howell}, J.}, \bibinfo{author}{{Hung}, T.},
  \bibinfo{author}{{Huppenkothen}, D.}, \bibinfo{author}{{Imel}, D.},
  \bibinfo{author}{{Ip}, W.H.}, \bibinfo{author}{{Ivezi{\'c}}, {\v{Z}}.},
  \bibinfo{author}{{Jackson}, E.}, \bibinfo{author}{{Jones}, L.},
  \bibinfo{author}{{Juric}, M.}, \bibinfo{author}{{Kasliwal}, M.M.},
  \bibinfo{author}{{Kaspi}, S.}, \bibinfo{author}{{Kaye}, S.},
  \bibinfo{author}{{Kelley}, M.S.P.}, \bibinfo{author}{{Kowalski}, M.},
  \bibinfo{author}{{Kramer}, E.}, \bibinfo{author}{{Kupfer}, T.},
  \bibinfo{author}{{Landry}, W.}, \bibinfo{author}{{Laher}, R.R.},
  \bibinfo{author}{{Lee}, C.D.}, \bibinfo{author}{{Lin}, H.W.},
  \bibinfo{author}{{Lin}, Z.Y.}, \bibinfo{author}{{Lunnan}, R.},
  \bibinfo{author}{{Giomi}, M.}, \bibinfo{author}{{Mahabal}, A.},
  \bibinfo{author}{{Mao}, P.}, \bibinfo{author}{{Miller}, A.A.},
  \bibinfo{author}{{Monkewitz}, S.}, \bibinfo{author}{{Murphy}, P.},
  \bibinfo{author}{{Ngeow}, C.C.}, \bibinfo{author}{{Nordin}, J.},
  \bibinfo{author}{{Nugent}, P.}, \bibinfo{author}{{Ofek}, E.},
  \bibinfo{author}{{Patterson}, M.T.}, \bibinfo{author}{{Penprase}, B.},
  \bibinfo{author}{{Porter}, M.}, \bibinfo{author}{{Rauch}, L.},
  \bibinfo{author}{{Rebbapragada}, U.}, \bibinfo{author}{{Reiley}, D.},
  \bibinfo{author}{{Rigault}, M.}, \bibinfo{author}{{Rodriguez}, H.},
  \bibinfo{author}{{van Roestel}, J.}, \bibinfo{author}{{Rusholme}, B.},
  \bibinfo{author}{{van Santen}, J.}, \bibinfo{author}{{Schulze}, S.},
  \bibinfo{author}{{Shupe}, D.L.}, \bibinfo{author}{{Singer}, L.P.},
  \bibinfo{author}{{Soumagnac}, M.T.}, \bibinfo{author}{{Stein}, R.},
  \bibinfo{author}{{Surace}, J.}, \bibinfo{author}{{Sollerman}, J.},
  \bibinfo{author}{{Szkody}, P.}, \bibinfo{author}{{Taddia}, F.},
  \bibinfo{author}{{Terek}, S.}, \bibinfo{author}{{Van Sistine}, A.},
  \bibinfo{author}{{van Velzen}, S.}, \bibinfo{author}{{Vestrand}, W.T.},
  \bibinfo{author}{{Walters}, R.}, \bibinfo{author}{{Ward}, C.},
  \bibinfo{author}{{Ye}, Q.Z.}, \bibinfo{author}{{Yu}, P.C.},
  \bibinfo{author}{{Yan}, L.}, \bibinfo{author}{{Zolkower}, J.},
  \bibinfo{year}{2019}.
\newblock \bibinfo{title}{{The Zwicky Transient Facility: System Overview,
  Performance, and First Results}}.
\newblock \bibinfo{journal}{Publications of the Astronomical Society of the
  Pacific} \bibinfo{volume}{131}, \bibinfo{pages}{018002}.
\newblock \DOIprefix\doi{10.1088/1538-3873/aaecbe},
  \href{http://arxiv.org/abs/1902.01932}{{\tt arXiv:1902.01932}}.
\bibitem[{{Bowell} et~al.(1989){Bowell}, {Hapke}, {Domingue}, {Lumme},
  {Peltoniemi} and {Harris}}]{1989aste.conf..524B}
\bibinfo{author}{{Bowell}, E.}, \bibinfo{author}{{Hapke}, B.},
  \bibinfo{author}{{Domingue}, D.}, \bibinfo{author}{{Lumme}, K.},
  \bibinfo{author}{{Peltoniemi}, J.}, \bibinfo{author}{{Harris}, A.W.},
  \bibinfo{year}{1989}.
\newblock \bibinfo{title}{{Application of photometric models to asteroids.}},
  in: \bibinfo{editor}{{Binzel}, R.P.}, \bibinfo{editor}{{Gehrels}, T.},
  \bibinfo{editor}{{Matthews}, M.S.} (Eds.), \bibinfo{booktitle}{Asteroids II},
  pp. \bibinfo{pages}{524--556}.
\bibitem[{{Bowell} and {Lumme}(1979)}]{1979aste.book..132B}
\bibinfo{author}{{Bowell}, E.}, \bibinfo{author}{{Lumme}, K.},
  \bibinfo{year}{1979}.
\newblock \bibinfo{title}{{Colorimetry and magnitudes of asteroids.}}
\bibitem[{{Collier Cameron} et~al.(2006){Collier Cameron}, {Pollacco},
  {Street}, {Lister}, {West}, {Wilson}, {Pont}, {Christian}, {Clarkson},
  {Enoch}, {Evans}, {Fitzsimmons}, {Haswell}, {Hellier}, {Hodgkin}, {Horne},
  {Irwin}, {Kane}, {Keenan}, {Norton}, {Parley}, {Osborne}, {Ryans}, {Skillen}
  and {Wheatley}}]{2006MNRAS.373..799C}
\bibinfo{author}{{Collier Cameron}, A.}, \bibinfo{author}{{Pollacco}, D.},
  \bibinfo{author}{{Street}, R.A.}, \bibinfo{author}{{Lister}, T.A.},
  \bibinfo{author}{{West}, R.G.}, \bibinfo{author}{{Wilson}, D.M.},
  \bibinfo{author}{{Pont}, F.}, \bibinfo{author}{{Christian}, D.J.},
  \bibinfo{author}{{Clarkson}, W.I.}, \bibinfo{author}{{Enoch}, B.},
  \bibinfo{author}{{Evans}, A.}, \bibinfo{author}{{Fitzsimmons}, A.},
  \bibinfo{author}{{Haswell}, C.A.}, \bibinfo{author}{{Hellier}, C.},
  \bibinfo{author}{{Hodgkin}, S.T.}, \bibinfo{author}{{Horne}, K.},
  \bibinfo{author}{{Irwin}, J.}, \bibinfo{author}{{Kane}, S.R.},
  \bibinfo{author}{{Keenan}, F.P.}, \bibinfo{author}{{Norton}, A.J.},
  \bibinfo{author}{{Parley}, N.R.}, \bibinfo{author}{{Osborne}, J.},
  \bibinfo{author}{{Ryans}, R.}, \bibinfo{author}{{Skillen}, I.},
  \bibinfo{author}{{Wheatley}, P.J.}, \bibinfo{year}{2006}.
\newblock \bibinfo{title}{{A fast hybrid algorithm for exoplanetary transit
  searches}}.
\newblock \bibinfo{journal}{Monthly Notices of the Royal Astronomical Society}
  \bibinfo{volume}{373}, \bibinfo{pages}{799--810}.
\newblock \DOIprefix\doi{10.1111/j.1365-2966.2006.11074.x},
  \href{http://arxiv.org/abs/astro-ph/0609418}{{\tt arXiv:astro-ph/0609418}}.
\bibitem[{{Dymock} and {Miles}(2009)}]{2009JBAA..119..149D}
\bibinfo{author}{{Dymock}, R.}, \bibinfo{author}{{Miles}, R.},
  \bibinfo{year}{2009}.
\newblock \bibinfo{title}{{A method for determining the V magnitude of
  asteroids from CCD images}}.
\newblock \bibinfo{journal}{Journal of the British Astronomical Association}
  \bibinfo{volume}{119}, \bibinfo{pages}{149--156}.
\newblock \href{http://arxiv.org/abs/1006.4017}{{\tt arXiv:1006.4017}}.
\bibitem[{{Hanu{\v{s}}} et~al.(2016){Hanu{\v{s}}}, {{\v{D}}urech},
  {Oszkiewicz}, {Behrend}, {Carry}, {Delbo}, {Adam}, {Afonina}, {Anquetin},
  {Antonini}, {Arnold}, {Audejean}, {Aurard}, {Bachschmidt}, {Baduel},
  {Barbotin}, {Barroy}, {Baudouin}, {Berard}, {Berger}, {Bernasconi}, {Bosch},
  {Bouley}, {Bozhinova}, {Brinsfield}, {Brunetto}, {Canaud}, {Caron},
  {Carrier}, {Casalnuovo}, {Casulli}, {Cerda}, {Chalamet}, {Charbonnel},
  {Chinaglia}, {Cikota}, {Colas}, {Coliac}, {Collet}, {Coloma}, {Conjat},
  {Conseil}, {Costa}, {Crippa}, {Cristofanelli}, {Damerdji}, {Deback{\`e}re},
  {Decock}, {D{\'e}hais}, {D{\'e}l{\'e}age}, {Delmelle}, {Demeautis},
  {Dr{\'o}{\.z}d{\.z}}, {Dubos}, {Dulcamara}, {Dumont}, {Durkee}, {Dymock},
  {Escalante del Valle}, {Esseiva}, {Esseiva}, {Esteban}, {Fauchez},
  {Fauerbach}, {Fauvaud}, {Fauvaud}, {Forn{\'e}}, {Fournel}, {Fradet},
  {Garlitz}, {Gerteis}, {Gillier}, {Gillon}, {Giraud}, {Godard}, {Goncalves},
  {Hamanowa}, {Hamanowa}, {Hay}, {Hellmich}, {Heterier}, {Higgins}, {Hirsch},
  {Hodosan}, {Hren}, {Hygate}, {Innocent}, {Jacquinot}, {Jawahar}, {Jehin},
  {Jerosimic}, {Klotz}, {Koff}, {Korlevic}, {Kosturkiewicz}, {Krafft},
  {Krugly}, {Kugel}, {Labrevoir}, {Lecacheux}, {Lehk{\'y}}, {Leroy},
  {Lesquerbault}, {Lopez-Gonzales}, {Lutz}, {Mallecot}, {Manfroid}, {Manzini},
  {Marciniak}, {Martin}, {Modave}, {Montaigut}, {Montier}, {Morelle}, {Morton},
  {Mottola}, {Naves}, {Nomen}, {Oey}, {Og{\l}oza}, {Paiella}, {Pallares},
  {Peyrot}, {Pilcher}, {Pirenne}, {Piron}, {Poli{\'n}ska}, {Polotto}, {Poncy},
  {Previt}, {Reignier}, {Renauld}, {Ricci}, {Richard}, {Rinner}, {Risoldi},
  {Robilliard}, {Romeuf}, {Rousseau}, {Roy}, {Ruthroff}, {Salom}, {Salvador},
  {Sanchez}, {Santana-Ros}, {Scholz}, {S{\'e}n{\'e}}, {Skiff}, {Sobkowiak},
  {Sogorb}, {Sold{\'a}n}, {Spiridakis}, {Splanska}, {Sposetti}, {Starkey},
  {Stephens}, {Stiepen}, {Stoss}, {Strajnic}, {Teng}, {Tumolo}, {Vagnozzi},
  {Vanoutryve}, {Vugnon}, {Warner}, {Waucomont}, {Wertz}, {Winiarski} and
  {Wolf}}]{2016A&A...586A.108H}
\bibinfo{author}{{Hanu{\v{s}}}, J.}, \bibinfo{author}{{{\v{D}}urech}, J.},
  \bibinfo{author}{{Oszkiewicz}, D.A.}, \bibinfo{author}{{Behrend}, R.},
  \bibinfo{author}{{Carry}, B.}, \bibinfo{author}{{Delbo}, M.},
  \bibinfo{author}{{Adam}, O.}, \bibinfo{author}{{Afonina}, V.},
  \bibinfo{author}{{Anquetin}, R.}, \bibinfo{author}{{Antonini}, P.},
  \bibinfo{author}{{Arnold}, L.}, \bibinfo{author}{{Audejean}, M.},
  \bibinfo{author}{{Aurard}, P.}, \bibinfo{author}{{Bachschmidt}, M.},
  \bibinfo{author}{{Baduel}, B.}, \bibinfo{author}{{Barbotin}, E.},
  \bibinfo{author}{{Barroy}, P.}, \bibinfo{author}{{Baudouin}, P.},
  \bibinfo{author}{{Berard}, L.}, \bibinfo{author}{{Berger}, N.},
  \bibinfo{author}{{Bernasconi}, L.}, \bibinfo{author}{{Bosch}, J.G.},
  \bibinfo{author}{{Bouley}, S.}, \bibinfo{author}{{Bozhinova}, I.},
  \bibinfo{author}{{Brinsfield}, J.}, \bibinfo{author}{{Brunetto}, L.},
  \bibinfo{author}{{Canaud}, G.}, \bibinfo{author}{{Caron}, J.},
  \bibinfo{author}{{Carrier}, F.}, \bibinfo{author}{{Casalnuovo}, G.},
  \bibinfo{author}{{Casulli}, S.}, \bibinfo{author}{{Cerda}, M.},
  \bibinfo{author}{{Chalamet}, L.}, \bibinfo{author}{{Charbonnel}, S.},
  \bibinfo{author}{{Chinaglia}, B.}, \bibinfo{author}{{Cikota}, A.},
  \bibinfo{author}{{Colas}, F.}, \bibinfo{author}{{Coliac}, J.F.},
  \bibinfo{author}{{Collet}, A.}, \bibinfo{author}{{Coloma}, J.},
  \bibinfo{author}{{Conjat}, M.}, \bibinfo{author}{{Conseil}, E.},
  \bibinfo{author}{{Costa}, R.}, \bibinfo{author}{{Crippa}, R.},
  \bibinfo{author}{{Cristofanelli}, M.}, \bibinfo{author}{{Damerdji}, Y.},
  \bibinfo{author}{{Deback{\`e}re}, A.}, \bibinfo{author}{{Decock}, A.},
  \bibinfo{author}{{D{\'e}hais}, Q.}, \bibinfo{author}{{D{\'e}l{\'e}age}, T.},
  \bibinfo{author}{{Delmelle}, S.}, \bibinfo{author}{{Demeautis}, C.},
  \bibinfo{author}{{Dr{\'o}{\.z}d{\.z}}, M.}, \bibinfo{author}{{Dubos}, G.},
  \bibinfo{author}{{Dulcamara}, T.}, \bibinfo{author}{{Dumont}, M.},
  \bibinfo{author}{{Durkee}, R.}, \bibinfo{author}{{Dymock}, R.},
  \bibinfo{author}{{Escalante del Valle}, A.}, \bibinfo{author}{{Esseiva}, N.},
  \bibinfo{author}{{Esseiva}, R.}, \bibinfo{author}{{Esteban}, M.},
  \bibinfo{author}{{Fauchez}, T.}, \bibinfo{author}{{Fauerbach}, M.},
  \bibinfo{author}{{Fauvaud}, M.}, \bibinfo{author}{{Fauvaud}, S.},
  \bibinfo{author}{{Forn{\'e}}, E.}, \bibinfo{author}{{Fournel}, C.},
  \bibinfo{author}{{Fradet}, D.}, \bibinfo{author}{{Garlitz}, J.},
  \bibinfo{author}{{Gerteis}, O.}, \bibinfo{author}{{Gillier}, C.},
  \bibinfo{author}{{Gillon}, M.}, \bibinfo{author}{{Giraud}, R.},
  \bibinfo{author}{{Godard}, J.P.}, \bibinfo{author}{{Goncalves}, R.},
  \bibinfo{author}{{Hamanowa}, H.}, \bibinfo{author}{{Hamanowa}, H.},
  \bibinfo{author}{{Hay}, K.}, \bibinfo{author}{{Hellmich}, S.},
  \bibinfo{author}{{Heterier}, S.}, \bibinfo{author}{{Higgins}, D.},
  \bibinfo{author}{{Hirsch}, R.}, \bibinfo{author}{{Hodosan}, G.},
  \bibinfo{author}{{Hren}, M.}, \bibinfo{author}{{Hygate}, A.},
  \bibinfo{author}{{Innocent}, N.}, \bibinfo{author}{{Jacquinot}, H.},
  \bibinfo{author}{{Jawahar}, S.}, \bibinfo{author}{{Jehin}, E.},
  \bibinfo{author}{{Jerosimic}, L.}, \bibinfo{author}{{Klotz}, A.},
  \bibinfo{author}{{Koff}, W.}, \bibinfo{author}{{Korlevic}, P.},
  \bibinfo{author}{{Kosturkiewicz}, E.}, \bibinfo{author}{{Krafft}, P.},
  \bibinfo{author}{{Krugly}, Y.}, \bibinfo{author}{{Kugel}, F.},
  \bibinfo{author}{{Labrevoir}, O.}, \bibinfo{author}{{Lecacheux}, J.},
  \bibinfo{author}{{Lehk{\'y}}, M.}, \bibinfo{author}{{Leroy}, A.},
  \bibinfo{author}{{Lesquerbault}, B.}, \bibinfo{author}{{Lopez-Gonzales},
  M.J.}, \bibinfo{author}{{Lutz}, M.}, \bibinfo{author}{{Mallecot}, B.},
  \bibinfo{author}{{Manfroid}, J.}, \bibinfo{author}{{Manzini}, F.},
  \bibinfo{author}{{Marciniak}, A.}, \bibinfo{author}{{Martin}, A.},
  \bibinfo{author}{{Modave}, B.}, \bibinfo{author}{{Montaigut}, R.},
  \bibinfo{author}{{Montier}, J.}, \bibinfo{author}{{Morelle}, E.},
  \bibinfo{author}{{Morton}, B.}, \bibinfo{author}{{Mottola}, S.},
  \bibinfo{author}{{Naves}, R.}, \bibinfo{author}{{Nomen}, J.},
  \bibinfo{author}{{Oey}, J.}, \bibinfo{author}{{Og{\l}oza}, W.},
  \bibinfo{author}{{Paiella}, M.}, \bibinfo{author}{{Pallares}, H.},
  \bibinfo{author}{{Peyrot}, A.}, \bibinfo{author}{{Pilcher}, F.},
  \bibinfo{author}{{Pirenne}, J.F.}, \bibinfo{author}{{Piron}, P.},
  \bibinfo{author}{{Poli{\'n}ska}, M.}, \bibinfo{author}{{Polotto}, M.},
  \bibinfo{author}{{Poncy}, R.}, \bibinfo{author}{{Previt}, J.P.},
  \bibinfo{author}{{Reignier}, F.}, \bibinfo{author}{{Renauld}, D.},
  \bibinfo{author}{{Ricci}, D.}, \bibinfo{author}{{Richard}, F.},
  \bibinfo{author}{{Rinner}, C.}, \bibinfo{author}{{Risoldi}, V.},
  \bibinfo{author}{{Robilliard}, D.}, \bibinfo{author}{{Romeuf}, D.},
  \bibinfo{author}{{Rousseau}, G.}, \bibinfo{author}{{Roy}, R.},
  \bibinfo{author}{{Ruthroff}, J.}, \bibinfo{author}{{Salom}, P.A.},
  \bibinfo{author}{{Salvador}, L.}, \bibinfo{author}{{Sanchez}, S.},
  \bibinfo{author}{{Santana-Ros}, T.}, \bibinfo{author}{{Scholz}, A.},
  \bibinfo{author}{{S{\'e}n{\'e}}, G.}, \bibinfo{author}{{Skiff}, B.},
  \bibinfo{author}{{Sobkowiak}, K.}, \bibinfo{author}{{Sogorb}, P.},
  \bibinfo{author}{{Sold{\'a}n}, F.}, \bibinfo{author}{{Spiridakis}, A.},
  \bibinfo{author}{{Splanska}, E.}, \bibinfo{author}{{Sposetti}, S.},
  \bibinfo{author}{{Starkey}, D.}, \bibinfo{author}{{Stephens}, R.},
  \bibinfo{author}{{Stiepen}, A.}, \bibinfo{author}{{Stoss}, R.},
  \bibinfo{author}{{Strajnic}, J.}, \bibinfo{author}{{Teng}, J.P.},
  \bibinfo{author}{{Tumolo}, G.}, \bibinfo{author}{{Vagnozzi}, A.},
  \bibinfo{author}{{Vanoutryve}, B.}, \bibinfo{author}{{Vugnon}, J.M.},
  \bibinfo{author}{{Warner}, B.D.}, \bibinfo{author}{{Waucomont}, M.},
  \bibinfo{author}{{Wertz}, O.}, \bibinfo{author}{{Winiarski}, M.},
  \bibinfo{author}{{Wolf}, M.}, \bibinfo{year}{2016}.
\newblock \bibinfo{title}{{New and updated convex shape models of asteroids
  based on optical data from a large collaboration network}}.
\newblock \bibinfo{journal}{Astronomy and Astrophysics} \bibinfo{volume}{586},
  \bibinfo{pages}{A108}.
\newblock \DOIprefix\doi{10.1051/0004-6361/201527441},
  \href{http://arxiv.org/abs/1510.07422}{{\tt arXiv:1510.07422}}.
\bibitem[{{Hapke}(2012)}]{2012Icar..221.1079H}
\bibinfo{author}{{Hapke}, B.}, \bibinfo{year}{2012}.
\newblock \bibinfo{title}{{Bidirectional reflectance spectroscopy 7. The single
  particle phase function hockey stick relation}}.
\newblock \bibinfo{journal}{Icarus} \bibinfo{volume}{221},
  \bibinfo{pages}{1079--1083}.
\newblock \DOIprefix\doi{10.1016/j.icarus.2012.10.022}.
\bibitem[{{Jewitt}(2013)}]{2013AJ....145..133J}
\bibinfo{author}{{Jewitt}, D.}, \bibinfo{year}{2013}.
\newblock \bibinfo{title}{{Properties of Near-Sun Asteroids}}.
\newblock \bibinfo{journal}{The Astronomical Journal} \bibinfo{volume}{145},
  \bibinfo{pages}{133}.
\newblock \DOIprefix\doi{10.1088/0004-6256/145/5/133},
  \href{http://arxiv.org/abs/1303.2415}{{\tt arXiv:1303.2415}}.
\bibitem[{{Jewitt} and {Hsieh}(2006)}]{2006AJ....132.1624J}
\bibinfo{author}{{Jewitt}, D.}, \bibinfo{author}{{Hsieh}, H.},
  \bibinfo{year}{2006}.
\newblock \bibinfo{title}{{Physical Observations of 2005 UD: A Mini-Phaethon}}.
\newblock \bibinfo{journal}{The Astronomical Journal} \bibinfo{volume}{132},
  \bibinfo{pages}{1624--1629}.
\newblock \DOIprefix\doi{10.1086/507483}.
\bibitem[{{Kasuga} and {Jewitt}(2008)}]{2008AJ....136..881K}
\bibinfo{author}{{Kasuga}, T.}, \bibinfo{author}{{Jewitt}, D.},
  \bibinfo{year}{2008}.
\newblock \bibinfo{title}{{Observations of 1999 YC and the Breakup of the
  Geminid Stream Parent}}.
\newblock \bibinfo{journal}{The Astronomical Journal} \bibinfo{volume}{136},
  \bibinfo{pages}{881--889}.
\newblock \DOIprefix\doi{10.1088/0004-6256/136/2/881},
  \href{http://arxiv.org/abs/0805.2636}{{\tt arXiv:0805.2636}}.
\bibitem[{{Kim} et~al.(2018){Kim}, {Lee}, {Lee}, {Kim}, {Yoshida}, {Bartczak},
  {Dudzi{\'n}ski}, {Park}, {Choi}, {Moon}, {Yim}, {Choi}, {Choi}, {Yoon},
  {Serebryanskiy}, {Krugov}, {Reva}, {Ergashev}, {Burkhonov}, {Ehgamberdiev},
  {Turayev}, {Lin}, {Arai}, {Ohtsuka}, {Ito}, {Urakawa} and
  {Ishiguro}}]{2018A&A...619A.123K}
\bibinfo{author}{{Kim}, M.J.}, \bibinfo{author}{{Lee}, H.J.},
  \bibinfo{author}{{Lee}, S.M.}, \bibinfo{author}{{Kim}, D.H.},
  \bibinfo{author}{{Yoshida}, F.}, \bibinfo{author}{{Bartczak}, P.},
  \bibinfo{author}{{Dudzi{\'n}ski}, G.}, \bibinfo{author}{{Park}, J.},
  \bibinfo{author}{{Choi}, Y.J.}, \bibinfo{author}{{Moon}, H.K.},
  \bibinfo{author}{{Yim}, H.S.}, \bibinfo{author}{{Choi}, J.},
  \bibinfo{author}{{Choi}, E.J.}, \bibinfo{author}{{Yoon}, J.N.},
  \bibinfo{author}{{Serebryanskiy}, A.}, \bibinfo{author}{{Krugov}, M.},
  \bibinfo{author}{{Reva}, I.}, \bibinfo{author}{{Ergashev}, K.E.},
  \bibinfo{author}{{Burkhonov}, O.}, \bibinfo{author}{{Ehgamberdiev}, S.A.},
  \bibinfo{author}{{Turayev}, Y.}, \bibinfo{author}{{Lin}, Z.Y.},
  \bibinfo{author}{{Arai}, T.}, \bibinfo{author}{{Ohtsuka}, K.},
  \bibinfo{author}{{Ito}, T.}, \bibinfo{author}{{Urakawa}, S.},
  \bibinfo{author}{{Ishiguro}, M.}, \bibinfo{year}{2018}.
\newblock \bibinfo{title}{{Optical observations of NEA 3200 Phaethon (1983 TB)
  during the 2017 apparition}}.
\newblock \bibinfo{journal}{Astronomy and Astrophysics} \bibinfo{volume}{619},
  \bibinfo{pages}{A123}.
\newblock \DOIprefix\doi{10.1051/0004-6361/201833593},
  \href{http://arxiv.org/abs/1809.05900}{{\tt arXiv:1809.05900}}.
\bibitem[{{Kinoshita} et~al.(2007){Kinoshita}, {Ohtsuka}, {Sekiguchi},
  {Watanabe}, {Ito}, {Arakida}, {Kasuga}, {Miyasaka}, {Nakamura} and
  {Lin}}]{2007A&A...466.1153K}
\bibinfo{author}{{Kinoshita}, D.}, \bibinfo{author}{{Ohtsuka}, K.},
  \bibinfo{author}{{Sekiguchi}, T.}, \bibinfo{author}{{Watanabe}, J.},
  \bibinfo{author}{{Ito}, T.}, \bibinfo{author}{{Arakida}, H.},
  \bibinfo{author}{{Kasuga}, T.}, \bibinfo{author}{{Miyasaka}, S.},
  \bibinfo{author}{{Nakamura}, R.}, \bibinfo{author}{{Lin}, H.C.},
  \bibinfo{year}{2007}.
\newblock \bibinfo{title}{{Surface heterogeneity of 2005 UD from photometric
  observations}}.
\newblock \bibinfo{journal}{Astronomy and Astrophysics} \bibinfo{volume}{466},
  \bibinfo{pages}{1153--1158}.
\newblock \DOIprefix\doi{10.1051/0004-6361:20066276}.
\bibitem[{{Krugly} et~al.(2019){Krugly}, {Belskaya}, {Mykhailova}, {Donchev},
  {Inasaridze}, {Sergeyev}, {Slyusarev}, {Shevchenko}, {Chiorny}, {Rumyantsev},
  {Novichonok}, {Ayvazian}, {Kapanadze}, {Kvaratskhelia}, {Bonev}, {Borisov},
  {Molotov} and {Voropaev}}]{2019EPSC...13.1989K}
\bibinfo{author}{{Krugly}, Y.}, \bibinfo{author}{{Belskaya}, I.N.},
  \bibinfo{author}{{Mykhailova}, S.S.}, \bibinfo{author}{{Donchev}, Z.},
  \bibinfo{author}{{Inasaridze}, R.Y.}, \bibinfo{author}{{Sergeyev}, A.V.},
  \bibinfo{author}{{Slyusarev}, I.G.}, \bibinfo{author}{{Shevchenko}, V.G.},
  \bibinfo{author}{{Chiorny}, V.G.}, \bibinfo{author}{{Rumyantsev}, V.V.},
  \bibinfo{author}{{Novichonok}, A.O.}, \bibinfo{author}{{Ayvazian}, V.},
  \bibinfo{author}{{Kapanadze}, G.}, \bibinfo{author}{{Kvaratskhelia}, O.I.},
  \bibinfo{author}{{Bonev}, T.}, \bibinfo{author}{{Borisov}, G.},
  \bibinfo{author}{{Molotov}, I.E.}, \bibinfo{author}{{Voropaev}, V.A.},
  \bibinfo{year}{2019}.
\newblock \bibinfo{title}{{Photometry and polarimetry of near-Earth asteroids
  (3200) Phaethon and (155140) 2005 UD}}, in: \bibinfo{booktitle}{EPSC-DPS
  Joint Meeting 2019}, pp. \bibinfo{pages}{EPSC--DPS2019--1989}.
\bibitem[{{Krugly} et~al.(2002){Krugly}, {Belskaya}, {Shevchenko}, {Chiorny},
  {Velichko}, {Mottola}, {Erikson}, {Hahn}, {Nathues}, {Neukum}, {Gaftonyuk}
  and {Dotto}}]{2002Icar..158..294K}
\bibinfo{author}{{Krugly}, Y.N.}, \bibinfo{author}{{Belskaya}, I.N.},
  \bibinfo{author}{{Shevchenko}, V.G.}, \bibinfo{author}{{Chiorny}, V.G.},
  \bibinfo{author}{{Velichko}, F.P.}, \bibinfo{author}{{Mottola}, S.},
  \bibinfo{author}{{Erikson}, A.}, \bibinfo{author}{{Hahn}, G.},
  \bibinfo{author}{{Nathues}, A.}, \bibinfo{author}{{Neukum}, G.},
  \bibinfo{author}{{Gaftonyuk}, N.M.}, \bibinfo{author}{{Dotto}, E.},
  \bibinfo{year}{2002}.
\newblock \bibinfo{title}{{The Near-Earth Objects Follow-up Program. IV. CCD
  Photometry in 1996-1999}}.
\newblock \bibinfo{journal}{Icarus} \bibinfo{volume}{158},
  \bibinfo{pages}{294--304}.
\newblock \DOIprefix\doi{10.1006/icar.2002.6884}.
\bibitem[{{Laher} et~al.(2014){Laher}, {Surace}, {Grillmair}, {Ofek},
  {Levitan}, {Sesar}, {van Eyken}, {Law}, {Helou}, {Hamam}, {Masci},
  {Mattingly}, {Jackson}, {Hacopeans}, {Mi}, {Groom}, {Teplitz}, {Desai},
  {Hale}, {Smith}, {Walters}, {Quimby}, {Kasliwal}, {Horesh}, {Bellm},
  {Barlow}, {Waszczak}, {Prince} and {Kulkarni}}]{2014PASP..126..674L}
\bibinfo{author}{{Laher}, R.R.}, \bibinfo{author}{{Surace}, J.},
  \bibinfo{author}{{Grillmair}, C.J.}, \bibinfo{author}{{Ofek}, E.O.},
  \bibinfo{author}{{Levitan}, D.}, \bibinfo{author}{{Sesar}, B.},
  \bibinfo{author}{{van Eyken}, J.C.}, \bibinfo{author}{{Law}, N.M.},
  \bibinfo{author}{{Helou}, G.}, \bibinfo{author}{{Hamam}, N.},
  \bibinfo{author}{{Masci}, F.J.}, \bibinfo{author}{{Mattingly}, S.},
  \bibinfo{author}{{Jackson}, E.}, \bibinfo{author}{{Hacopeans}, E.},
  \bibinfo{author}{{Mi}, W.}, \bibinfo{author}{{Groom}, S.},
  \bibinfo{author}{{Teplitz}, H.}, \bibinfo{author}{{Desai}, V.},
  \bibinfo{author}{{Hale}, D.}, \bibinfo{author}{{Smith}, R.},
  \bibinfo{author}{{Walters}, R.}, \bibinfo{author}{{Quimby}, R.},
  \bibinfo{author}{{Kasliwal}, M.}, \bibinfo{author}{{Horesh}, A.},
  \bibinfo{author}{{Bellm}, E.}, \bibinfo{author}{{Barlow}, T.},
  \bibinfo{author}{{Waszczak}, A.}, \bibinfo{author}{{Prince}, T.A.},
  \bibinfo{author}{{Kulkarni}, S.R.}, \bibinfo{year}{2014}.
\newblock \bibinfo{title}{{IPAC Image Processing and Data Archiving for the
  Palomar Transient Factory}}.
\newblock \bibinfo{journal}{Publications of the Astronomical Society of the
  Pacific} \bibinfo{volume}{126}, \bibinfo{pages}{674}.
\newblock \DOIprefix\doi{10.1086/677351},
  \href{http://arxiv.org/abs/1404.1953}{{\tt arXiv:1404.1953}}.
\bibitem[{{Masiero} et~al.(2019){Masiero}, {Wright} and
  {Mainzer}}]{2019AJ....158...97M}
\bibinfo{author}{{Masiero}, J.R.}, \bibinfo{author}{{Wright}, E.L.},
  \bibinfo{author}{{Mainzer}, A.K.}, \bibinfo{year}{2019}.
\newblock \bibinfo{title}{{Thermophysical Modeling of NEOWISE Observations of
  DESTINY$^{+}$ Targets Phaethon and 2005 UD}}.
\newblock \bibinfo{journal}{The Astronomical Journal} \bibinfo{volume}{158},
  \bibinfo{pages}{97}.
\newblock \DOIprefix\doi{10.3847/1538-3881/ab31a6},
  \href{http://arxiv.org/abs/1907.04518}{{\tt arXiv:1907.04518}}.
\bibitem[{{Muinonen} et~al.(2010){Muinonen}, {Belskaya}, {Cellino},
  {Delb{\`o}}, {Levasseur-Regourd}, {Penttil{\"a}} and
  {Tedesco}}]{2010Icar..209..542M}
\bibinfo{author}{{Muinonen}, K.}, \bibinfo{author}{{Belskaya}, I.N.},
  \bibinfo{author}{{Cellino}, A.}, \bibinfo{author}{{Delb{\`o}}, M.},
  \bibinfo{author}{{Levasseur-Regourd}, A.C.}, \bibinfo{author}{{Penttil{\"a}},
  A.}, \bibinfo{author}{{Tedesco}, E.F.}, \bibinfo{year}{2010}.
\newblock \bibinfo{title}{{A three-parameter magnitude phase function for
  asteroids}}.
\newblock \bibinfo{journal}{Icarus} \bibinfo{volume}{209},
  \bibinfo{pages}{542--555}.
\newblock \DOIprefix\doi{10.1016/j.icarus.2010.04.003}.
\bibitem[{{Muinonen} et~al.(2015){Muinonen}, {Wilkman}, {Cellino}, {Wang} and
  {Wang}}]{2015P&SS..118..227M}
\bibinfo{author}{{Muinonen}, K.}, \bibinfo{author}{{Wilkman}, O.},
  \bibinfo{author}{{Cellino}, A.}, \bibinfo{author}{{Wang}, X.},
  \bibinfo{author}{{Wang}, Y.}, \bibinfo{year}{2015}.
\newblock \bibinfo{title}{{Asteroid lightcurve inversion with Lommel-Seeliger
  ellipsoids}}.
\newblock \bibinfo{journal}{Planetary and Space Science} \bibinfo{volume}{118},
  \bibinfo{pages}{227--241}.
\newblock \DOIprefix\doi{10.1016/j.pss.2015.09.005}.
\bibitem[{{Ofek} et~al.(2012){Ofek}, {Laher}, {Law}, {Surace}, {Levitan},
  {Sesar}, {Horesh}, {Poznanski}, {van Eyken}, {Kulkarni}, {Nugent},
  {Zolkower}, {Walters}, {Sullivan}, {Ag{\"u}eros}, {Bildsten}, {Bloom},
  {Cenko}, {Gal-Yam}, {Grillmair}, {Helou}, {Kasliwal} and
  {Quimby}}]{2012PASP..124...62O}
\bibinfo{author}{{Ofek}, E.O.}, \bibinfo{author}{{Laher}, R.},
  \bibinfo{author}{{Law}, N.}, \bibinfo{author}{{Surace}, J.},
  \bibinfo{author}{{Levitan}, D.}, \bibinfo{author}{{Sesar}, B.},
  \bibinfo{author}{{Horesh}, A.}, \bibinfo{author}{{Poznanski}, D.},
  \bibinfo{author}{{van Eyken}, J.C.}, \bibinfo{author}{{Kulkarni}, S.R.},
  \bibinfo{author}{{Nugent}, P.}, \bibinfo{author}{{Zolkower}, J.},
  \bibinfo{author}{{Walters}, R.}, \bibinfo{author}{{Sullivan}, M.},
  \bibinfo{author}{{Ag{\"u}eros}, M.}, \bibinfo{author}{{Bildsten}, L.},
  \bibinfo{author}{{Bloom}, J.}, \bibinfo{author}{{Cenko}, S.B.},
  \bibinfo{author}{{Gal-Yam}, A.}, \bibinfo{author}{{Grillmair}, C.},
  \bibinfo{author}{{Helou}, G.}, \bibinfo{author}{{Kasliwal}, M.M.},
  \bibinfo{author}{{Quimby}, R.}, \bibinfo{year}{2012}.
\newblock \bibinfo{title}{{The Palomar Transient Factory Photometric
  Calibration}}.
\newblock \bibinfo{journal}{Publications of the Astronomical Society of the
  Pacific} \bibinfo{volume}{124}, \bibinfo{pages}{62}.
\newblock \DOIprefix\doi{10.1086/664065},
  \href{http://arxiv.org/abs/1112.4851}{{\tt arXiv:1112.4851}}.
\bibitem[{{Ohtsuka} et~al.(2006){Ohtsuka}, {Sekiguchi}, {Kinoshita},
  {Watanabe}, {Ito}, {Arakida} and {Kasuga}}]{2006A&A...450L..25O}
\bibinfo{author}{{Ohtsuka}, K.}, \bibinfo{author}{{Sekiguchi}, T.},
  \bibinfo{author}{{Kinoshita}, D.}, \bibinfo{author}{{Watanabe}, J.I.},
  \bibinfo{author}{{Ito}, T.}, \bibinfo{author}{{Arakida}, H.},
  \bibinfo{author}{{Kasuga}, T.}, \bibinfo{year}{2006}.
\newblock \bibinfo{title}{{Apollo asteroid 2005 UD: split nucleus of (3200)
  Phaethon?}}
\newblock \bibinfo{journal}{Astronomy and Astrophysics} \bibinfo{volume}{450},
  \bibinfo{pages}{L25--L28}.
\newblock \DOIprefix\doi{10.1051/0004-6361:200600022}.
\bibitem[{{Shevchenko} et~al.(2016){Shevchenko}, {Belskaya}, {Muinonen},
  {Penttil{\"a}}, {Krugly}, {Velichko}, {Chiorny}, {Slyusarev}, {Gaftonyuk} and
  {Tereschenko}}]{2016P&SS..123..101S}
\bibinfo{author}{{Shevchenko}, V.G.}, \bibinfo{author}{{Belskaya}, I.N.},
  \bibinfo{author}{{Muinonen}, K.}, \bibinfo{author}{{Penttil{\"a}}, A.},
  \bibinfo{author}{{Krugly}, Y.N.}, \bibinfo{author}{{Velichko}, F.P.},
  \bibinfo{author}{{Chiorny}, V.G.}, \bibinfo{author}{{Slyusarev}, I.G.},
  \bibinfo{author}{{Gaftonyuk}, N.M.}, \bibinfo{author}{{Tereschenko}, I.A.},
  \bibinfo{year}{2016}.
\newblock \bibinfo{title}{{Asteroid observations at low phase angles. IV.
  Average parameters for the new H, G$_{1}$, G$_{2}$ magnitude system}}.
\newblock \bibinfo{journal}{Planetary and Space Science} \bibinfo{volume}{123},
  \bibinfo{pages}{101--116}.
\newblock \DOIprefix\doi{10.1016/j.pss.2015.11.007}.
\bibitem[{{Slivan}(2002)}]{2002Natur.419...49S}
\bibinfo{author}{{Slivan}, S.M.}, \bibinfo{year}{2002}.
\newblock \bibinfo{title}{{Spin vector alignment of Koronis family asteroids}}.
\newblock \bibinfo{journal}{Nature} \bibinfo{volume}{419},
  \bibinfo{pages}{49--51}.
\newblock \DOIprefix\doi{10.1038/nature00993}.
\bibitem[{{Tamuz} et~al.(2005){Tamuz}, {Mazeh} and
  {Zucker}}]{2005MNRAS.356.1466T}
\bibinfo{author}{{Tamuz}, O.}, \bibinfo{author}{{Mazeh}, T.},
  \bibinfo{author}{{Zucker}, S.}, \bibinfo{year}{2005}.
\newblock \bibinfo{title}{{Correcting systematic effects in a large set of
  photometric light curves}}.
\newblock \bibinfo{journal}{Monthly Notices of the Royal Astronomical Society}
  \bibinfo{volume}{356}, \bibinfo{pages}{1466--1470}.
\newblock \DOIprefix\doi{10.1111/j.1365-2966.2004.08585.x},
  \href{http://arxiv.org/abs/astro-ph/0502056}{{\tt arXiv:astro-ph/0502056}}.
\bibitem[{{Wang} et~al.(2013){Wang}, {Gu}, {Collier Cameron}, {Tan}, {Hui},
  {Kwok}, {Yeung} and {Leung}}]{2013RAA....13..593W}
\bibinfo{author}{{Wang}, X.B.}, \bibinfo{author}{{Gu}, S.H.},
  \bibinfo{author}{{Collier Cameron}, A.}, \bibinfo{author}{{Tan}, H.B.},
  \bibinfo{author}{{Hui}, H.K.}, \bibinfo{author}{{Kwok}, C.T.},
  \bibinfo{author}{{Yeung}, B.}, \bibinfo{author}{{Leung}, K.C.},
  \bibinfo{year}{2013}.
\newblock \bibinfo{title}{{The refined physical parameters of transiting
  exoplanet system HAT-P-24}}.
\newblock \bibinfo{journal}{Research in Astronomy and Astrophysics}
  \bibinfo{volume}{13}, \bibinfo{pages}{593--603}.
\newblock \DOIprefix\doi{10.1088/1674-4527/13/5/010}.
\bibitem[{{Warner} and {Stephens}(2019)}]{2019MPBu...46..144W}
\bibinfo{author}{{Warner}, B.D.}, \bibinfo{author}{{Stephens}, R.D.},
  \bibinfo{year}{2019}.
\newblock \bibinfo{title}{{Near-Earth Asteroid Lightcurve Analysis at the
  Center for Solar System Studies: 2018 September-December}}.
\newblock \bibinfo{journal}{Minor Planet Bulletin} \bibinfo{volume}{46},
  \bibinfo{pages}{144--152}.

\end{thebibliography}





\end{document}